\documentclass{aa}  

\usepackage{graphicx}
\pdfoutput=1
\usepackage{txfonts}
\usepackage{natbib}
\usepackage{placeins}
\usepackage{multicol}
\newcommand{\fe}    {[Fe\,{\sc ii}]}

\newcommand{\hdos}  {H$_2$}

\begin{document}

   \title{NIR spectroscopic survey of protostellar jets in the star-forming region IC 1396N\thanks{Based on observations
made with the Italian Telescopio Nazionale Galileo (TNG) operated
on the island of La Palma by the Fundaci\'{o}n Galileo Galilei of the INAF (Istituto Nazionale
di Astrofisica) at the Spanish Observatorio del Roque de los Muchachos of the
Instituto de Astrofisica de Canarias (time awarded by IAC, programme CAT\_32/2007, P.\ I.\  M.\ Beltr\'{a}n).}
\fnmsep\thanks{Table 3, Table 4, and Table 5 are only available in electronic form
at the CDS via anonymous ftp to cdsarc.u-strasbg.fr (130.79.128.5)
or via http://cdsweb.u-strasbg.fr/cgi-bin/qcat?J/A+A/.}}

\titlerunning{IC 1396N: spectroscopy of H$_2$ jets}

\author{
F. Massi\inst{1}
\and
R. L\'opez\inst{2}
\and
M. T. Beltr\'an\inst{1}
\and
R. Estalella\inst{2}
\and
J. M. Girart\inst{3,4}
}

\institute{
INAF - Osservatorio Astrofisico di Arcetri, Largo E. Fermi 5, I-50125 Firenze,
Italy.
\and
Departament de F\'{\i}sica Qu\`antica i Astrof\'{\i}sica,
Institut de Ci\`encies del Cosmos, Universitat de Barcelona, IEEC-UB,
Mart\'{i} i Franqu\`es, 1, E-08028 Barcelona, Spain.
\and
Institut de Ci\`encies de l'Espai (ICE), CSIC, 
Carrer de Can Magrans, s/n, E-08193 Cerdanyola del Vall\`es, Catalonia, Spain.
\and
Institut d'Estudis Espacials de Catalunya (IEEC), E-08034 Barcelona, Catalonia, Spain
}

   \date{received ; accepted }

 
  \abstract
{The bright-rimmed cloud \object{IC 1396N}, associated with an intermediate-mass star-forming region,
hosts a number of CO, molecular hydrogen, and Herbig-Haro (HHs) outflows powered by  a set of 
millimetre compact sources.
}
{The aim of this work is to  characterise the kinematics and physical conditions of the \hdos{}
emission features spread throughout the \object{IC 1396N} region. The features appear as chains of knots with a jet-like morphology and trace different
\hdos{} outflows. We also obtain further information about (and an identification of) the driving sources.  
}
{Low-resolution, long-slit near-infrared spectra were acquired with the NICS camera at the TNG telescope,
using grisms KB ($R \sim 1200$), HK, and JH ($R \sim 500$). Several slit pointings and position angles were used throughout the \object{IC 1396N} 
region in order to sample a number of the \hdos{} knots that were previously detected in deep H$_2$ $2.12$ $\mu$m
images.
}
{The knots exhibit rich ro-vibrational spectra of H$_2$, consistent with shock-excited excitation, from which 
radial velocities and relevant physical conditions of the \object{IC 1396N} \hdos{}  outflows were derived.
These also allowed estimating extinction ranges towards several features. \fe\ emission was only detected
towards a few knots that also display unusually high H$_2$ 1--0 S(3)/S(1) flux ratios. 
The obtained radial velocities confirm that most of the outflows are close to the plane of the sky.
Nearby knots in the same chain often display different radial velocities, both blue-shifted and red-shifted, which we interpret as due
to ubiquitous jet precession in the driving sources or the development of oblique shocks. 
One of the chains 
(strand A, i. e. knots A1 to A15) 
appears as a set
of features trailing a leading bow-shock structure consistent with the results of 3D magneto-hydrodynamical
models. The sides of the leading bow shock (A15) exhibit different radial velocities. We discuss possible explanations. 
Our data cannot confirm whether strands A and B have both originated in the intermediate mass young stellar
object \object{[BGE2002] BIMA 2} because a simple model of a precessing jet cannot account for their locations.
}
{Near-infrared spectroscopy has confirmed that most of the H$_2$ ro-vibrational emission in \object{IC 1396N} is shock-excited rather
than uv-excited in photon-dominated regions. It has shown a complex kinematical structure in most strands of emitting
knots as well.}


   \maketitle
%

\section{Introduction}

The bright-rimmed cloud \object{IC 1396N} (also known as BRC38; \citealt{1991ApJS...77...59S}), 
associated with the \ion{H}{ii} region \object{IC 1396}, is an interesting laboratory
for studying a range of star formation processes in progress. It exhibits a cometary structure, whose
bright rim faces the ionising star \object{HD 206267}  towards the south, and its tail points to the north.
\object{HD 206267}  is a multiple system with \object{HD 206267}  Aa (O5V+B0V) and Ab (O9V), and is located in 
the cluster \object{Trumpler 37} \citep{2020A&A...636A..28M} inside the \ion{H}{ii} region IC1396. The distance to the system estimated from GAIA
parallaxes is $1093^{+67}_{-59}$ pc \citep{2020A&A...636A..28M}. This value points to a slightly larger distance 
to \object{IC 1396N} than assumed in previous works (usually 750 pc). This is also confirmed by other authors
using GAIA data. Thus, we adopt here a revised distance of $910 \pm 49$ pc from the catalogue
of distances to molecular clouds of \cite{2020A&A...633A..51Z}, obtained by combining stellar photometric data and
GAIA parallaxes. Stellar optical polarisation
measurements show that the magnetic field in the cloud is almost aligned with the ionising 
radiation, forming an average angle of $\sim 20 \degr$ with it \citep{2018MNRAS.476.4782S}. The magnetic field is in fact roughly
parallel to the Galactic plane, as expected on a larger scale.

The head of the cloud is characterised by diffuse emission in the mid-infrared (MIR; see e.g. Fig.~5 of
\citealt{2009A&A...504...97B}), whereas the tail appears as a dark lane running from south to
north (see e.g. the $JHK$ image in Fig.~1 of \citealt{2009A&A...504...97B}). This morphology is
well reflected in millimetre (mm) line emission (\citealt{2001A&A...376..271C}; \citealt{2002aprm.conf..213S}). 
The star formation activity inside the cloud is clearly evident because
both HH objects are visible at optical wavelengths \citep{2003ApJ...593L..47R} and chains of knots
of H$_2$ $2.12$ $\mu$m line emission are detected in the near-infrared (NIR; \citealt{2001A&A...376..553N}; \citealt{2002aprm.conf..213S};
\citealt{2006A&A...449.1077C}; \citealt{2009A&A...504...97B}). Most of the H$_2$ knots originated in shocked gas in the outflows rather than in uv-excited fluorescence, as confirmed by their coincidence
with mm outflows (\citealt{2001A&A...376..271C}; \citealt{2002ApJ...573..246B}; \citealt{2012A&A...542L..26B}). 
The stellar population around the head was studied in the NIR, MIR, and in
X-rays to discuss possible scenarios
of triggered star formation (\citealt{2007ApJ...654..316G}; \citealt{2009A&A...504...97B};
\citealt{2010ApJ...717.1067C}). 

The H$_2$ knots towards the head of \object{IC 1396N} are associated with the intermediate-mass young stellar object (YSO)
\object{IRAS 21391+5802} (and other nearby YSOs; \citealt{2002ApJ...573..246B}). Other H$_2$ knots and the
Herbig-Haro object \object{HH 777} in the middle of the cloud are associated with the NIR source \object{HH 777}/IRS331
(\citealt{2003ApJ...593L..47R}; \citealt{2002ApJ...573..246B}), whereas most of the H$_2$ knots in
the tail of the cloud are associated with two mm compact sources (\citealt{2001A&A...376..271C}; 
\citealt{2012A&A...542L..26B}). In particular, \citet{2012A&A...542L..26B} discussed what was probably
the first documented occurrence of the interaction between outflows, causing a complex structure of
H$_2$ line emission. So far, only one other case of a possible outflow collision has been
reported (in \object{BHR 71}; see \citealt{2018AJ....156..239Z}).

The ro-vibrational line $v = 1 - 0$ $S(1)$ at $2.12$ $\mu$m of molecular hydrogen has proven to be
an efficient tracer of protostellar jets and outflows through narrow-band imaging. H$_2$ 
ro-vibrational emission spectra originate in
slow, non-dissociative C-shocks or in the post-shock regions of fast ($\ga 25 - 50$ km s$^{-1}$)
dissociative J-shocks (e.g. \citealt{1979ApJS...41..555H}; \citealt{1980ApJ...241L..47H};
\citealt{1989ApJ...342..306H};
\citealt{1996ApJ...456..611K}). H$_2$ line emission also occurs in photon-dominated regions (PDRs) 
due to fluorescent excitation
\citep{1987ApJ...322..412B}. In principle, radiative fluorescent emission can be distinguished from
thermal emission based on the ro-vibrational spectra if gas densities are $\la 10^{4}$ cm$^{-3}$,
whereas thermal emission becomes dominant at higher gas densities even with intense UV fields
\citep{1989ApJ...338..197S}. In this respect, higher-excitation ro-vibrational lines falling in the $J$ and $H$
bands can be instrumental in distinguishing the excitation mechanism
\citep{1987ApJ...322..412B}, which unfortunately happens to be a spectral region 
that is more extincted than $K$. Other interesting lines may occur in the $JH$ band, such as
the [\ion{Fe}{II}] lines originating from the upper level $3 d^{6} (^{5}D) 4s\, a^{4} D_{7/2}$
of ionised iron
at $1.6440$, $1.3209$, and $1.2570$ $\mu$m, which trace J-shocks \citep{1989ApJ...342..306H}. When these are present, they can be used to estimate extinction \citep{2016PASP..128g3001P}. 

With this in mind, we carried out a long-slit low-resolution ($R \sim 500$ to $1250$) survey of \object{IC 1396N} in the $JHK$ bands
to distinguish the different flows (and to distinguish between shock-excited and fluorescent-excited H$_2$ emission)
in the head of the cloud and identify new possible driving sources. 
\citet{2006A&A...449.1077C} have reported on long-slit low-resolution ($R \sim 500$) NIR spectra
of targets in \object{IC 1396N}, but they are limited to the two chains (or strands) of knots
called A and B towards the head of the cloud without information about the single knots
of each chain.

The paper is laid out as follows: the observations and data reduction
are described in Sect.~\ref{ODR}, the main results are described in Sect.~\ref{resu}
and are further discussed in Sect.~\ref{discu}. Sect.~\ref{conclu} summarises the main
findings, and some important issues are detailed in the appendix.

\section{Observations and data reduction}
\label{ODR}

\begin{figure}
\resizebox{\hsize}{!}{\includegraphics{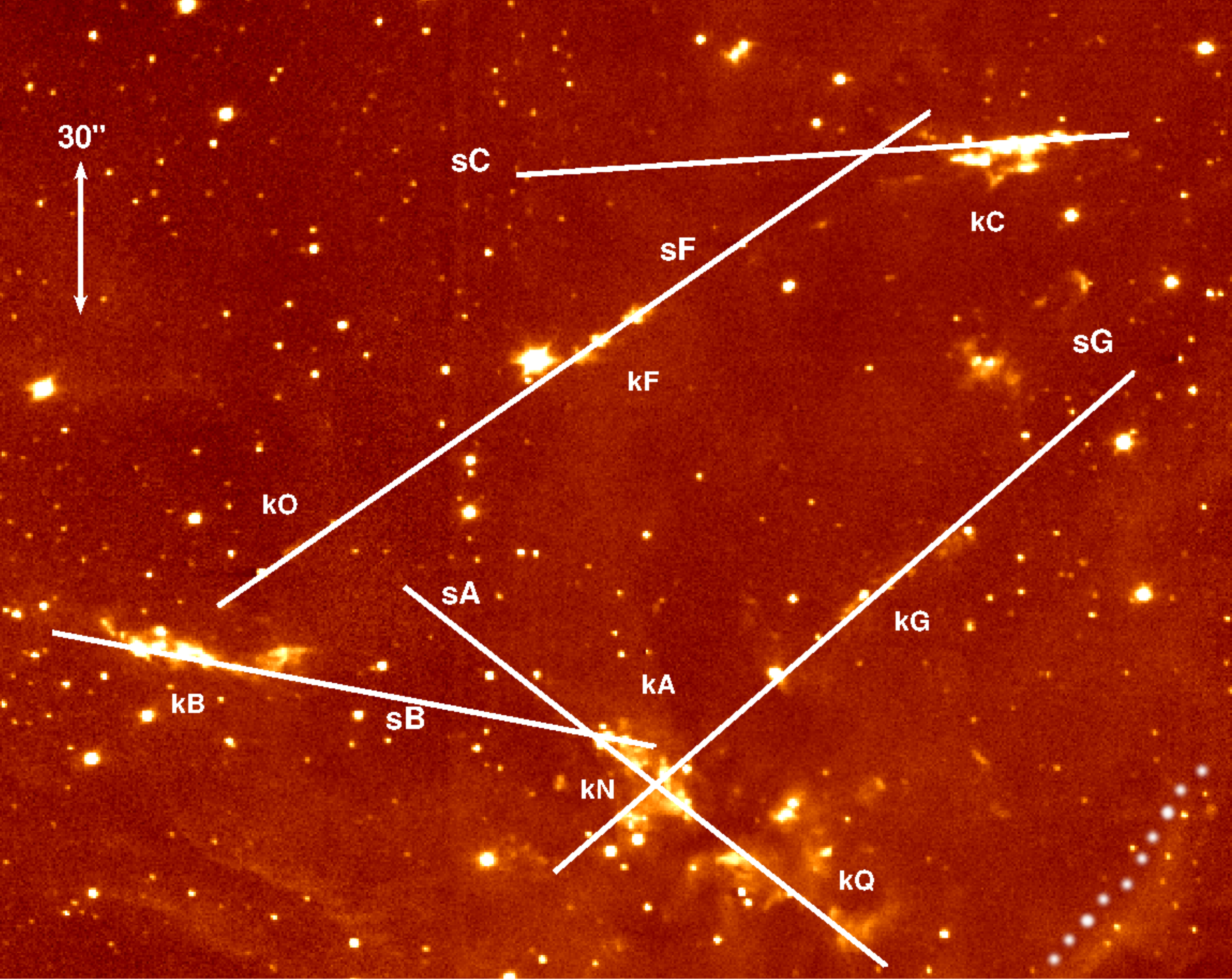}}
\caption{\label{fig:span:slits}
H2 (line plus continuum) image of \object{IC 1396N} from \citet{2009A&A...504...97B} overlaid with
the slit locations selected for this work (sX: slit; kX: knot chain). The dotted line
in the bottom right corner marks the photo-ionised strip encompassed by slit A. 
}
\end{figure}

Low-resolution long-slit spectra were obtained at several slit positions
(see Fig.~\ref{fig:span:slits}) with the NIR camera NICS \citep{2001A&A...378..722B}
at the 3.58~m Telescopio Nazionale Galileo (TNG)
on August 22, 23, and 24, 2007, using grisms KB (spanning the spectral range $1.95 - 2.34$ $\mu$m), HK
($1.40 - 2.50$ $\mu$m), and JH ($1.15 - 1.75$ $\mu$m), and a $1\arcsec$ wide slit,
achieving spectral resolutions of 1250, 500, and 500, respectively. The plate scale is $0.25\arcsec$/pix,
yielding a 
field of view (fov) of $4\farcm2 \times 4\farcm2$, whereas the slit length is $4\arcmin$.
For each integration, pointing
and slit position angle (PA) were selected so that several knots were encompassed by the
slit at the same time (see Fig.~\ref{fig:span:slits}). The slit was positioned on the plane of the sky by using reference stars.
PA and reference stars were chosen based on a careful analysis of the H$_2$ $2.12$ $\mu$m
image of \citet{2009A&A...504...97B}.
All observations were carried out as multiple AB cycles, taking two exposures with
the target shifted along the slit (i.e. in positions A and B). However, as several knots span 
the slit length in various positions, the slit was moved completely off-source in the B position.  
Spectra of two telluric standards, an A0 star (\object{HIP 109079}) and a G2V star (\object{HIP 110327}),
were obtained with one ABBA cycle each per grism at the
beginning, in the middle, and at the end of each night, with the slit roughly oriented at
the parallactic angle. On July 24, \object{HIP 109079} was imaged in various positions along the slit to
allow deriving a set of traces in different parts of the frame to use as references in the
subsequent extraction of the target spectra.
Details are given in Table~\ref{spect:tba}. Halogen and Ar lamp spectra
were obtained in the afternoon preceding each observing night
for flat-fielding and wavelength calibrating.

%
%
\begin{table*}[htb]
\tiny
\centering
\caption{\label{spect:tba}
Log of spectral observations.
}
\begin{tabular}{lllllll}
\hline
Date & Targeted & grism &        Slit           & Parallactic & Number of   & Integration time \\
     &   knots ($^{a}$) &       & position angle        & angle ($^{b}$)      & AB cycles ($^{c}
$) & single exposure \\
     &          &       &         ($\degr$)         &  ($\degr$)      &           & (s)
\\
\hline
22/08/2007 & \object{HIP 109079} & KB                     & 0  &  95       &  1 ABBA & 20 \\
           &    A      & KB                     & $52.5$ & 62     &  1      & 600 \\
           &    B      & KB                     & $80.5$ & 51     &  1      & 600 \\
           &    C      & KB                     & $94.5$ & 35      &  1      & 600 \\
           & \object{HIP 109079} & KB                     & 55 & 45      &  1 ABBA & 30 \\
           &    F      & KB                     & 125 & 158  &  2      & 900 \\
           &    G      & KB                     & 130 & 121  &  2      & 900 \\
           & \object{HIP 109079} & KB                     & 100 & 96   &  1 ABBA & 20 \\
23/08/2007 & \object{HIP 109079} & HK                     & $-85$ & $-87$ &  1 ABBA &  9 \\
           & \object{HIP 110327} & HK                     & $-95$ & $-95$     &  1 ABBA & 20 \\
           & C       & HK                     & $108.5$ & 64      &  4 & 900      \\
           & C       & JH                     & $108.5$ & 12     &  4 & 900      \\
           & \object{HIP 109079} & JH                     & 110     & 129      &  1 ABBA &  7 \\
           & \object{HIP 109079} & HK                     & 110     & 127      &  1 ABBA &  7 \\
           & \object{HIP 110327} & HK                     & 140     & 144      &  1 ABBA & 16 \\
           & \object{HIP 110327} & JH                     & 140     & 142     &  1 ABBA & 16 \\
           & G       & JH                     & 130     &  121        &  4 & 900      \\
           & \object{HIP 110327} & JH                     & 90 & 96     &  1 ABBA & 16 \\
           & \object{HIP 109079} & JH                     & 80 & 82     &  1 ABBA &  5 \\
24/08/2007 & \object{HIP 109079} & JH                     & 100 & 94     &  1 ABBA &  9 \\
           & \object{HIP 110327} & JH                     & 90 &  86    &  1 ABBA & 30 \\
           & AB      & JH                     & $79.5$ & 117  &  4 & 900      \\
           & \object{HIP 109079} & JH                     & 40 & 43     &  1 ABBA &  9 \\
           & \object{HIP 110327} & JH                     & 30 & 24 24  &  1 ABBA & 20 \\
           & F       & JH                     & 125 &  170       &  4 & 900      \\
           & A       & JH                     & $52,5$  & 114         &  3 & 900      \\
           & \object{HIP 109079} & JH                     & $-90$   & $-94$      &  1 ABBA &  9 \\
           & \object{HIP 110327} & JH                     & $-90$   & $-87$     &  1 ABBA & 35 \\
\hline
\end{tabular}
\tablefoot{
\tablefoottext{a}
{Following the notation used in \citet{2009A&A...504...97B}.}
\tablefoottext{b}
{$180\degr$ is sometimes added to the actual parallactic angle for comparison with the
slit position angle.}
\tablefoottext{c}
{Position B was still on-source for the telluric stars, but it was off-source for the science fields.}
}
\end{table*}

The spectra were reduced using standard  {\small IRAF} routines\footnote{{\small
IRAF} is distributed
by the National Optical Astronomy Observatories, which are operated by the
Association of Universities for Research in Astronomy, Inc., under cooperative
agreement with the National Science Foundation.}. For each night run, the halogen lamp frames
taken in the afternoon were averaged together and normalised row by row to a third-degree polynomial fit, as advised in the TNG web-page. The science frames
were corrected for flat field and for cross-talk using the software provided by the TNG staff,
and bad pixels were removed. For each AB cycle, the second frame was then subtracted from
the first frame. The 2D spectra were then straightened and wavelength calibrated using
the lines in the afternoon Ar lamp spectra. When the same target was acquired with more than
one AB cycle, we checked that the spatial profiles in a bright emission line were
consistent with each other. The subtracted 2D frames were then added together
(generally, no shift was needed in the
spatial direction to overlap the spectra). In a few cases only did  we find some
inconsistencies, and we did not combine the frames. These were clearly caused by small 
pointing differences.

There is usually some residual shift (a few \AA\ to a few dozen \AA)
along  the wavelength axis between the Ar lamp frames
and the target frames. This shift is variable with pointing location and slit position angle
(see \citealt{2008A&A...490.1079M}). We therefore re-aligned the science
and telluric standard frames taken during the same night by using sky emission lines as references.
Finally, we extracted 1D spectra for each identified knot and each telluric standard star in all
2D frames. For the standard stars, we were able to derive a suitable trace (i.e. the curve
describing the image of a point at a given slit position in a 2D frame) from the star itself. For each knot, however, we used
the trace derived from the stellar continuum spectrum as a reference that lay nearest to
the location of the knot in the frame out of the
set of observations made in the same night and through the same grism. This trace was then shifted
along the spatial direction
to the peak of the brightest emission line of the knot, and its width was reset accordingly. This was
done because no continuum emission is associated with the knots. Only a few emission lines exist,
which prevents us from deriving the traces from their own spectra. Background
subtraction was performed as well by choosing nearby spatial intervals on the same 2D spectrum
that were devoid of line
emission. This further corrected for residual sky emission that was still present in the subtracted frames.

%
%

Several knots were encompassed by the slit for each pointing, as described above.
These were easily identified on
the 2D spectra before extraction by using the following procedure. First, we
extracted the whole spatial profile throughout the slit length at the wavelength of a bright emission line.
For each 2D spectrum, we then rotated the H$_{2}$ image from \citet{2009A&A...504...97B} by the same position angle,
and we extracted the sum of four adjoining columns (the plate scale is the same as for the spectra; four columns correspond to $1\arcsec$, i.e. the slit width) passing through the approximate
slit centre location of the corresponding spectra. Shifting the set of summed columns along the perpendicular
direction by a few pixels soon resulted in spatial profiles that matched those obtained from the spectra
very well. This allowed us to identify the relevant knots.

The observed telluric standard stars
were used to correct the spectra for telluric absorption. Correction
spectra were obtained from the A0V star with {\it xtellcor}, which uses a high-resolution
spectrum of Vega to remove the intrinsic stellar
hydrogen lines, following \cite{2003PASP..115..389V}. Correction spectra were
also obtained from the G2V star following \cite{1996AJ....111..537M}, using a high-resolution spectrum of the Sun
to remove the intrinsic stellar metallic lines. By multiplying a correction and a
science spectrum, the effects of atmospheric transmission are removed provided the
two spectra were taken at a similar airmass and as close as possible in time. 

More details about the telluric correction are given in Appendix~\ref{tel:cor:ap}. 
As discussed there, we do not expect the uncertainties on the telluric absorption correction
to affect line ratios by more than $\sim 10$ \% in the clear part
of the observed spectral band (outside the yellow areas in Fig.~\ref{fig_atmo}). Lines coinciding with
telluric absorption bands of CO$_2$, O$_2$, or CH$_4$ may have larger errors, but these are probably still at the
$\sim 10$ \% level. However, \cite{2016PASP..128g3001P} reported that the flux of lines occurring in wavelength regions
characterised by crowded unsaturated telluric absorption bands is likely to be underestimated when measured in low-resolution spectra. In addition, the flux of lines falling in the wings of the strong telluric water absorption regions
should be regarded with caution.

\section{Results}
\label{resu}

\subsection{Line kinematics}
\label{l:k:s}

We used the molecular hydrogen lines extracted from the higher-resolution spectra
(grism KB) to derive some kinematical information. All lines appear unresolved in
the dispersion direction, therefore the emitting gas always spans a velocity range that is narrower than the
spectral resolution element ($R \sim 1250$, i.e.  $\Delta V \la 240$ km s$^{-1}$).
Nevertheless, radial velocities can be obtained with a better  accuracy than the spectral resolution.

We measured the wavelengths of the identified 1--0 S(0), S(1), and S(2) transitions (i.e.
the brightest detected lines) by fitting Gaussian distributions to the line profiles. We assumed rest (vacuum)
wavelengths of $2.223290$, $2.121834$, and $2.033758$ $\mu$m for the S(0), S(1), and S(2)
transition, respectively, from \cite{2019A&A...630A..58R}. We then refined the wavelength
calibration by measuring
the radial velocities of the OH sky lines in the same (unsubtracted) frames. We identified the OH lines falling in intervals
of $\sim 0.07$ $\mu$m centred around the rest wavelength of the three transitions and derived
a local linear correction in each of the three intervals (the frames were all previously
wavelength calibrated and straightened  with the lamp spectra).
The (vacuum) wavelengths of the reference sky lines were taken from \cite{2000A&A...354.1134R}.
The r.m.s. of the measured wavelengths minus the Rousselot
wavelengths of the OH lines were $\la 0.0001 - 0.0002$ $\mu$m (i.e. $\sim 15 - 30$
km s$^{-1}$), confirming that the observed
H$_2$ line wavelengths can be derived with an accuracy  better
than the spectral resolution. Finally, the differences between observed 
(lamp calibrated)
and theoretical H$_2$
wavelength after the OH-based correction were converted into radial velocities.

We note that the uncertainty on the radial velocities has two components. A systematic uncertainty
related to the wavelength calibration, which we assume to be equal to the r.m.s. of the observed
(lamp-calibrated)
minus theoretical wavelengths of the comparison  OH sky lines in each considered interval. And a random uncertainty, which we
assume to be equal to the line width divided by the signal-to-noise ratio of each H$_2$
feature. Thus, H$_2$ features originating from the three transitions we used may have different
systematic errors in the same frame. Conversely, H$_2$ features from different knots
in the same frame and originating from the same H$_2$ transition have the same systematic error
and can be compared to each other  with better accuracy.

The radial velocities we determined  are listed in Table~\ref{rad:vel}. Only those derived
from the H$_2$ 1--0 S(1) transition (i.e. the feature with the best 
signal-to-noise ratio, S/N) are
indicated. However, when the S(0) and S(2) transitions are also detected with good S/N, the
derived radial velocities are consistent with each other
within the errors, even more so when a shift is applied to the radial velocities obtained from the S(0) and S(2) transitions
so that they coincide with that derived from the S(1) transition
for the brightest knot in each frame to minimise the effects of the systematic component of the uncertainty.

%
%
\begin{table}
\tiny
\centering
\caption{\label{rad:vel}
Radial velocities (with respect to the local standard of rest) obtained from the
H$_2$ 1--0 S(1) line. The listed knots can be identified in Fig.~\protect\ref{find:chart}. 
}
\begin{tabular}{ll}
\hline
knot($^{a}$) & V$_{\rm LSR}$($^{b}$)\\
     & (km s$^{-1}$)\\
\hline
\multicolumn{2}{c}{frame A($^{c})$}\\
\hline
A1 & $3 \pm 3$ ($8.5$)\\
A4 & $8 \pm 3$ ($8.5$)\\
A8 & $-1 \pm 5$ ($8.5$)\\
A11 & $3 \pm 5$ ($8.5$)\\
A14 & $41 \pm 2$ ($8.5$)\\
A15 south & $50 \pm 2$ ($8.5$)\\
A16 & $8 \pm 1$ ($8.5$)\\
photo-ionised strip & $-17 \pm 11$ ($8.5$)\\
Q1 & $-18 \pm 27$ ($8.5$)\\
Q5 & $-15 \pm 8$ ($8.5$)\\
Q6 & $-15 \pm 12$ ($8.5$)\\
\hline
\multicolumn{2}{c}{frame B($^{c}$)}\\
\hline
A12 & $-36 \pm 4$ ($23$) \\
A15 north & $-34 \pm 4$ ($23$) \\
east of B2 & $5 \pm 5$ ($23$) \\
B3 & $-15 \pm 3$ ($23$) \\
between B3/B4 & $-17 \pm 4$ ($23$) \\
B4 & $-8 \pm 2$ ($23$) \\
B5 & $-10 \pm 2$ ($23$) \\
B7 & $8 \pm 7$ ($23$) \\
B8 & $11 \pm 1$ ($23$) \\
I3 & $0 \pm 14$ ($23$) \\
S2 & $15 \pm 21$ ($23$) \\
\hline
\multicolumn{2}{c}{frame C($^{c}$)}\\
\hline
C6 & $-47 \pm 4$ ($23$) \\
C8 & $-25 \pm 0.4$ ($23$) \\
unclassified east of C8/C10 & $-33 \pm 1$ ($23$) \\
C10 & $-43 \pm 1$ ($23$) \\
C14 & $-8 \pm 1$ ($23$) \\
unclassified north-east of C14 & $-45 \pm 1$ ($23$) \\
C16 & $-10 \pm 2$ ($23$) \\
\hline
\multicolumn{2}{c}{frame F($^{c}$)}\\
\hline
C1 & $-21 \pm 10$ ($28$) \\
F1 & $9 \pm 5$ ($28$) \\
F2 & $-11 \pm 3$ ($28$) \\
F4 & $-33 \pm 7$ ($28$) \\
F5 & $4 \pm 4$ ($28$) \\
O1 & $22 \pm 7$ ($28$) \\
\hline
\multicolumn{2}{c}{frame G($^{c}$)}\\
\hline
A6/A7 & $-10 \pm 1$ ($21$) \\
plateau A7 & $3 \pm 3$ ($21$) \\
G1 & $-7 \pm 3$ ($21$) \\
G2 & $2 \pm 2$ ($21$) \\
G3 & $40  \pm 1$ ($21$) \\
G4 & $-17 \pm 2$ ($21$) \\
G7 & $-31 \pm 5$ ($21$) \\
N1 & $22 \pm 6$ ($21$) \\
\hline
\end{tabular}
\tablefoot{
\tablefoottext{a}
{Following the notation used in \citet{2009A&A...504...97B}.}
\tablefoottext{b}
{the random error is estimated as the line width divided by line S/N; the
systematic component of uncertainty (due to the wavelength
calibration uncertainty) is indicated in brackets.}
\tablefoottext{c}
{The knots listed below were recorded simultaneously in the same frame.}
}
\end{table}
\addtocounter{table}{3}

\begin{figure}
\resizebox{\hsize}{!}{\includegraphics{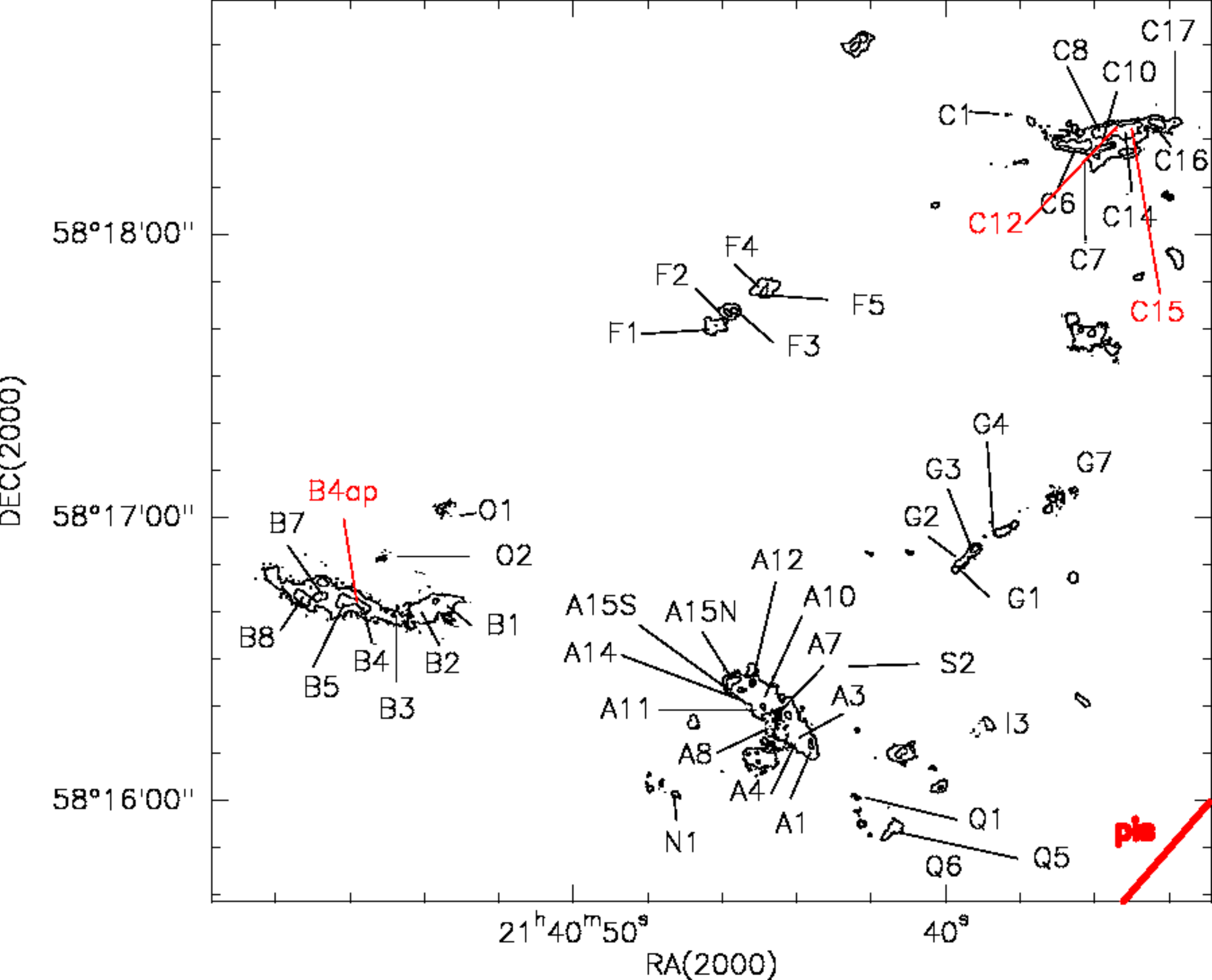}}
\caption{\label{find:chart}
Map of the H$_2$ emission features in \object{IC 1396N} 
from the H$_2$ $2.12$ $\mu$m
image of \citet{2009A&A...504...97B}. 
The knots discussed in the text are labelled.
The location of the photo-ionised strip (pis) is only indicative.
}
\end{figure}

One interesting result concerns 
knots A1 to A15 
(see Fig.~\ref{find:chart} to identify the emission features discussed
in the text). As can be deduced from Table~\ref{rad:vel}, the southern chain
of knots (A15 south, A14, A11) are red-shifted with velocities in the range 3 -- 50 km s$^{-1}$, whereas
the northern chain (A15 north, A12, A6/A7) are blue-shifted with velocities in the range $-10$ to
$-36$ km s$^{-1}$. This is fully consistent with the mm data from the CO(1--0) line (see Fig.~7
of \citealt{2009A&A...504...97B}), where blue-shifted ($-3.5$ to $-9.5$ km s$^{-1}$) and red-shifted ($3.5$ to $9.5$ km s$^{-1}$)
emission overlap towards the strand of 
knots A1--A15  
and split in the eastern tip, with blue-shifted emission in the north
and red-shifted emission in the south. This is evident with knot A15, a V-shaped feature that consists of
two peaks that we call A15 south and A15 north. These two peaks are clearly well separated in radial velocity
beyond the errors: 
A15 south is red-shifted and A15 north is blue-shifted. 
The adjoining knots A12 and A14 roughly
exhibit the same velocities, suggesting that A15 south and A14 constitute a separate system from A15 north and A12.
One possibility is that they are delineating a large bow shock moving eastward on the plane of the sky.

Knots C1--C17 
are also consistent with CO(1--0) emission (see Fig. 2 of \citealt{2012A&A...542L..26B}), which delineates a cavity
towards C with velocities ranging from $\sim -10$ to $\sim -40$ km s$^{-1}$
(the systemic velocity is $\sim 0$ km s$^{-1}$), the same as exhibited by the
knots of H$_{2}$ NIR emission. Interestingly, the chain composed of knots C6, C8, and C10, facing source C of
\citet{2012A&A...542L..26B}, clearly consists of the most blue-shifted knots beyond the errors.

CO(1--0) emission (see Fig.~1 of \citealt{2012A&A...542L..26B}) indicates that 
knots F1--F5 
should be red-shifted with radial velocities
in the range 0--20 km s$^{-1}$. Table~\ref{rad:vel} shows that F1 and F5 are consistent with the CO data.
Nevertheless, F2 and F4 are clearly blue-shifted compared to F1 and F5 beyond the errors. This may suggest that
knots F1--F5 trace an oblique shock
or possibly  jet precession.

For
knots B1--B11, 
two separate chains of blue-shifted (B3, B4, and B5) and red-shifted (B7 and B8) knots
are clearly identifiable. This requires further mm investigation to study the outflow structure more closely.
If associated with \object{[BGE2002] BIMA 2,} as strand A is thought to be as proposed by \citet{2009A&A...504...97B}, this would confirm jet precession
as suggested by its wiggled morphology.

%
%

The strand of knots G is likely to be associated with the NIR source 331 of \citet{2009A&A...504...97B} 
located south-east
of them. This source is also thought to originate HH 777, perpendicular to strand G, suggesting that it
is a double star \citep{2003ApJ...593L..47R}. The farthest knots
from 331, G4 and G7, are blue-shifted, while the nearest knots become progressively more red-shifted from G1
to G3, with G3 clearly red-shifted above the errors compared to all the others. This again suggests a
complex structure of the flow.

One of the slits perpendicularly encompassed  a strip of emission running south-east north-west and located south-west 
of strand A, marking the PDR
that borders the molecular cloud (hereafter the ''photo-ionised strip''; it is barely visible in the bottom right
corner of Fig.~\ref{fig:span:slits}; see also Fig.~\ref{find:chart}). Interestingly, this is clearly
blue-shifted compared to 
knots A1--A15, 
indicating that the heated gas in the PDR is expanding outward of the cloud.

\subsection{Line fluxes}
\label{li:fl:ds}

The knots in strand C 
have the largest spectral range coverage. They were
observed through both grisms HK
and JH. Grism KB spectra were meant to mainly study kinematics and include several knots of 
strands A, B, C, F, and G
(see Table~\ref{rad:vel}). Finally, we used grism JH to search for [\ion{Fe}{II}] emission lines as well,
observing knots in strands A, B, C, F, and G.

All detected spectra are composed of ro-vibrational lines of H$_2$, and in only a few cases, 
of the forbidden lines [\ion{Fe}{II}] at $1.257$, $1.320$, and $1.644$
$\mu$m. The measured fluxes are listed in Table~3 and Table~4, normalised for each knot to the flux of the
H$_2$ 1--0 S(1) line at $2.1218$ $\mu$m in the HK and KB grism bands, and to the H$_2$ 1--0 S(7) line at $1.7480$ $\mu$m
in the $JH$ band. We adopted the knot designation used in  \citet{2009A&A...504...97B} with a few additions as some of the line emission
is from features that were not classified in \citet{2009A&A...504...97B}. Table~5 provides an updated photometry of the H$_2$ 1--0 S(1)
line emission for the knots detected in our spectra based on the narrow-band image of \citet{2009A&A...504...97B}. This
photometry can be used to calibrate the fluxes in the HK and KB grism bands (and in the $JH$ band when the corresponding $HK$ band
spectrum is available as well). The photometry has been revised to include the knots
that were previously not classified. We note that all the main physical properties of the knots can be derived from the normalised fluxes, as
discussed in the next section.

\subsection{Line ratios}
\label{li:ra:ds}

The ratio of the H$_2$ 1--0 S(1) line at $2.1218$ $\mu$m to the 2--1 S(1) line at $2.2477$ $\mu$m ($R_{21}$) is
often used to distinguish between shock-excited and UV-fluorescent H$_2$ emission. Where the 2--1
line is detected (knots C, A, and G), $R_{21} \sim 10$. In other cases, lower limits ($R_{21} > 3-5$) are found (see
Tables~3 and 4). $R_{21} \ga 10$ is usually considered an indication of shock excitation, although
quiescent dense gas near to intense UV sources can yield high $R_{21}$ values as well \citep{1989ApJ...338..197S}. In our
case, most of the knots are associated with outflows, which confirms the shock nature. For the
photo-ionised strip, we obtain $R_{21} \sim 2.1$. This value is typical of radiative
fluorescent emission at densities $\la 10^{4}$ cm$^{-3}$ \citep{1989ApJ...338..197S}. Even an
extinction $A_V \sim 20$ would cause $R_{21}$ to be underestimated by only $\sim 20$ \%.
Another notably difference between the photo-ionised strip and the other knots is the ratio of the
H$_2$ 1--0 S(1) line to the H$_2$ 1--0 S(0) line at $2.2233$ $\mu$m. Most knots exhibit a value of
$\sim 3-5$, whereas the strip exhibits a ratio $\sim 1$. Both the 1--0 S(0) and the 1--0  S(2)
lines display fluxes comparable to that of the 1--0 S(1) line. The main ionising source originating
in the PDR is the O5V star HD2006267. Assuming it is at a distance of $\sim 13$ pc from the cloud,
which is the projected separation between star and cloud at a distance of 910 pc,
we can estimate that the intensity in the boundary radiation field  is $\sim 100$ times the
average value in the solar neighbourhood (in the units used by \citealt{1987ApJ...322..412B}, with the
parameters given for an O5V star by \citealt{1996ApJ...460..914V} and either the relation
scaled to the flux at 1000 \AA\ given by \citealt{1987ApJ...322..412B} or equation A6 of
\citealt{1989ApJ...338..197S}). 
With this UV field,
the models of \citet{1987ApJ...322..412B} predict intense fluorescently excited H$_2$ 1--0 S(0) and S(2)
lines, with ratios S(0)/S(1) and S(2)/S(1) $\sim 0.5 - 0.6$ for molecular gas densities in the
range $10^{2}$--$10^{4}$ cm$^{-3}$. The signal-to-noise ratio of the photo-ionised strip
spectrum is too low to allow us to
carry out a more detailed analysis, other than noting that it clearly looks different from
the spectra of the other knots and is consistent with fluorescent excitation.

\begin{figure}
\resizebox{\hsize}{!}{\includegraphics{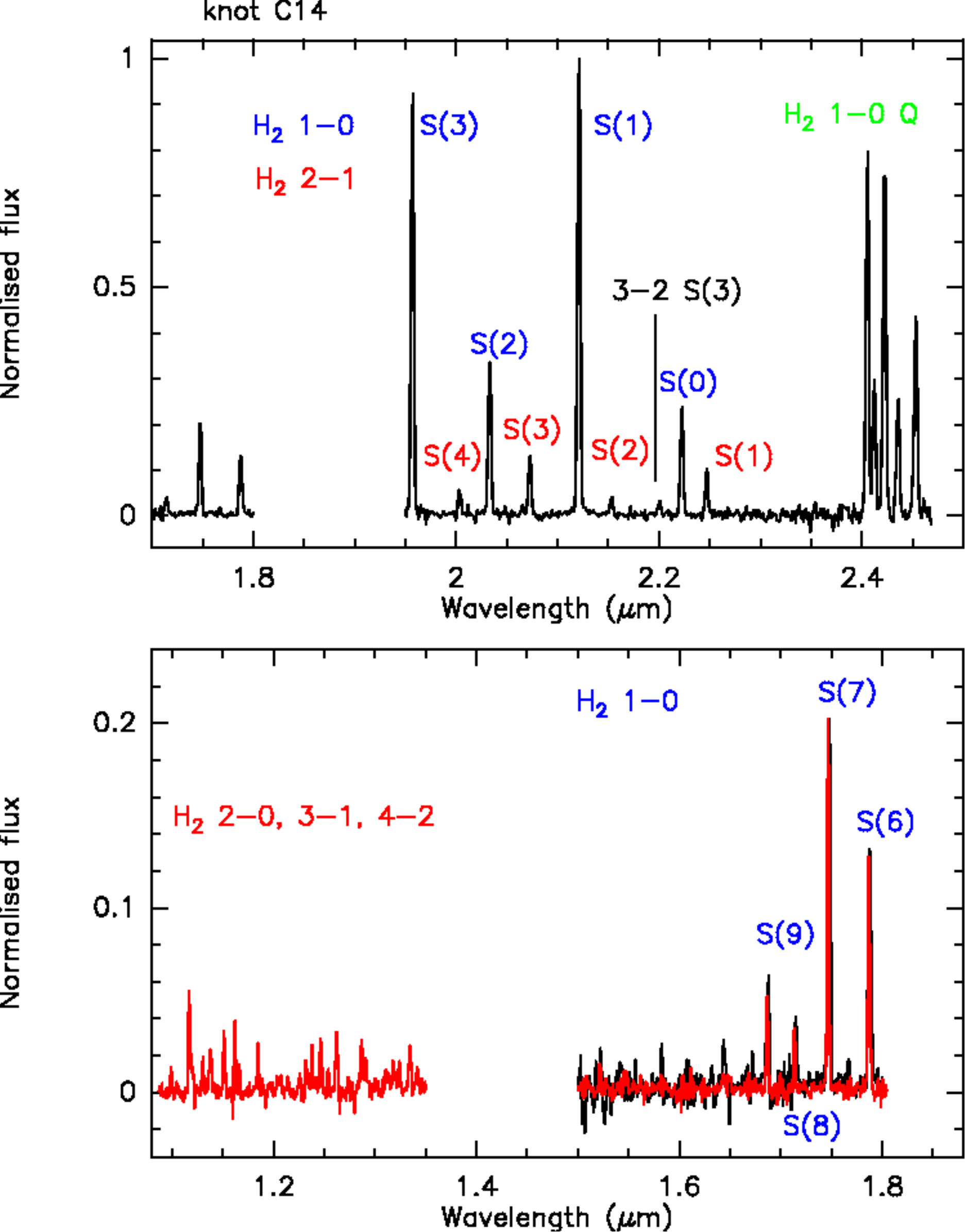}}
\caption{\label{fig_C14}
$HK$ and $JH$ spectrum of knot C14. The flux scale in $HK$ (black) is normalised to
the peak flux density of the H$_2$ 1--0  S(1) emission, and 
the $JH$ spectrum (red) has been scaled such that line H$_2$ 1--0  S(7) exhibits the same
normalised flux in the $JH$ and $HK$ spectra. The identified molecular
hydrogen ro-vibrational lines are indicated.
}
\end{figure}

\begin{figure}
\resizebox{\hsize}{!}{\includegraphics{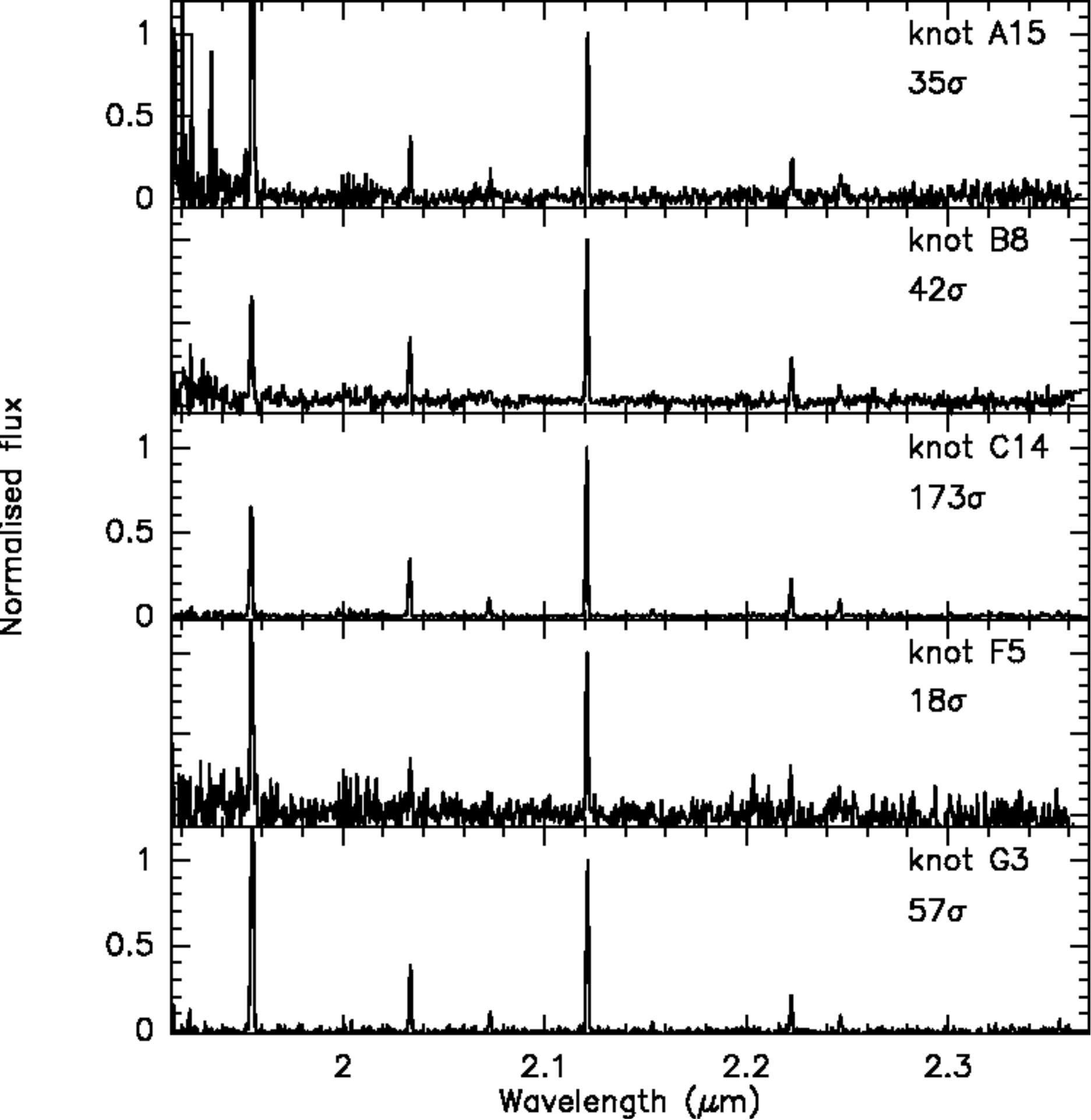}}
\caption{\label{fig:KB}
Comparison of the KB spectra of knots A15, B8, C14, F5, and G3. The ratio of the
H$_2$ 1--0  S(1) peak flux density and the r.m.s. in the adjacent spectral regions is
indicated in each panel.
}
\end{figure}

More accurate constraints on the physical conditions of the emitting gas can be obtained from the
normalised fluxes (Tables~3 and 4) where a large number of H$_2$ ro-vibrational lines are detected.
Knots C1--C17 
exhibit the richest spectra, with detected H$_2$ lines from $v=1-0, 2-1, 2-0, 3-1, 4-2$
transitions throughout $JHK$ (see Fig.~\ref{fig_C14}), partly due to the good S/N we achieved. 
The S/N 
depends not only on knot
brightness and exposure time, but also on how well the slit was positioned over the knots. Due
to the blind pointing, we were not able to optimise the slit positions straight on the knots.
This is easily seen in Fig~\ref{fig:KB}, which compares the KB spectra of knots
A15, B8, C14, F5, and
G3. Clearly, the better the S/N, the higher the number of detected transitions.
We used the H$_2$ 1--0 S(6), S(7), S(8), and S(9) lines at $1.7880$, $1.7480$, $1.7147$, and $1.6878$ $\mu$m,
respectively, to inter-calibrate the $JH$ and $HK$ spectra of 
knots C1--C17. 
The fluxes were weighted according to their
spectrophotometric errors. In a few cases, these four lines exhibit slightly different ratios between $JH$ and
$HK$, partly because S(6) and S(7) lie in a region in which atmospheric absorption is more
critical (see Fig.~\ref{fig_atmo}). However, this also indicates that the slit positions in $JH$ and $HK$ were slightly different.
We note again that the slit was set on the image field by using fixed position angles and reference stars (the knots are too
faint to set the slit directly). The complete spectrum of knot C14 is shown in Fig.~\ref{fig_C14}
as an example.

%
%
%
%

%
%

\begin{figure}
\resizebox{\hsize}{!}{\includegraphics{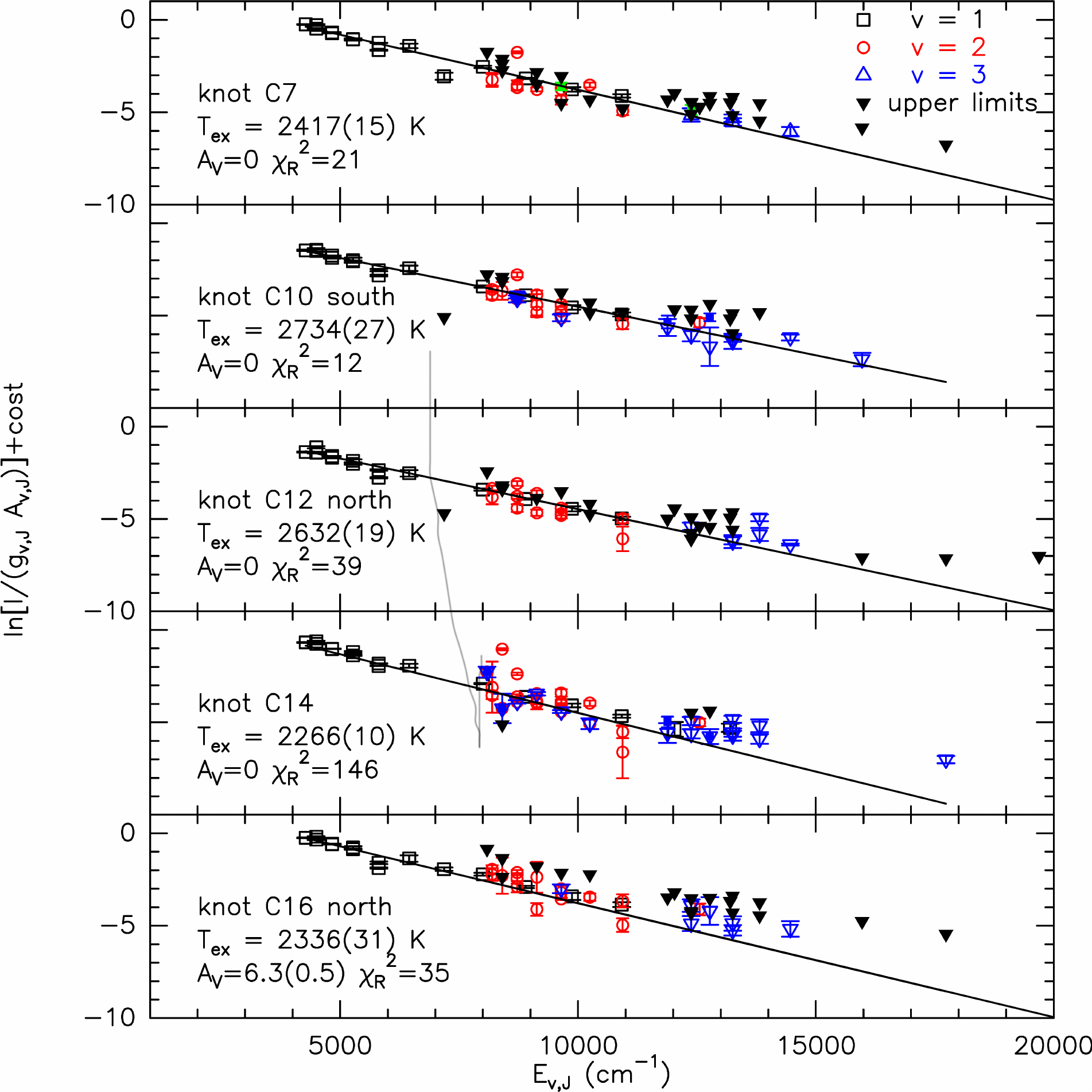}}
\caption{\label{fit:fig}
Ro-vibrational plots (Boltzmann plots) of the H$_2$ lines from 
knots C1--C17, 
with full $JHK$ coverage.
Open black squares mark transitions from the upper vibrational level $v=1$,
open red circles transitions from $v=2$, open blue triangles transitions from $v=3$,
and full black triangles upper limits for various undetected transitions. 
The straight lines mark the best linear fits
obtained for both the excitation temperature and the extinction
simultaneously. The
data displayed have been corrected for the extinction value providing the best fit.
 The obtained temperature, extinction, and reduced $\chi^{2}$ are also
indicated in each panel.
}
\end{figure}

In principle, line ratios of H$_2$ transitions originating from the same upper level can
be used to derive the extinction towards a knot. In our case 1--0 S($i$) and 1--0 Q($i+2$),
and 2--1 S($i$) and 2--0 S($i$) might be suitable. Unfortunately, each pair of lines
provides different results because either one of the transitions is faint or falls in
a region in which atmospheric correction is important, for instance, Q($i+2$), or the line ratio
is affected by the $JH$--$HK$ inter-calibration error. In addition, this computation requires that a knot
has at least been observed through the HK grism. The KB grism band encloses no suitable line pairs.
However, all H$_2$ detected lines can be used to simultaneously derive the excitation temperature
$T_{\rm ex}$ and the extinction $A_{V}$ minimising the uncertainty. Assuming that the gas
is thermalised, the points in a ro-vibrational diagram plotting $\ln[I/(g_{v,J}A_{v,J})]$
versus\ $E_{v,J}$ (i.e. a so-called  Boltzmann plot)  are aligned in a straight line provided the line flux $I$ has been
 corrected
for extinction. In our notation, $A_{v,J}$ is the ro-vibrational transition probability, $E_{v,J}$ is the energy of
the upper level, and $g_{v,J}$ is the statistical weight. As the line slope depends on $T_{\rm ex}$,
a simple linear fit, varying $A_{V}$,   simultaneously provides the best approximating
$T_{\rm ex}$ and $A_{V}$. By adopting the extinction law
of \citet{1985ApJ...288..618R} in the form prescribed by \citet{2005A&A...441..159N}, which is based on \citet{1989ESASP.290...93D},
 using the transition probabilities given in \citet{1977ApJS...35..281T} and deriving $E_{v,J}$
from \citet{huber}, we performed the fit for 
knots C1--C17 
with full $JHK$ coverage, using only the points
with S/N $> 3$. These fits are shown in Fig.~\ref{fit:fig}.

%
%

The obtained $T_{\rm ex}$ (2000--3000 K) is typical of H$_2$ emission of knots from Class 0 and Class I
sources (see e.g. \citealt{2006A&A...449.1077C}). To evaluate the reliability of the linear fits, we repeated
them by using the lines that fell in only one of the HK, JH, and KB grism bands. By limiting the fit to
the $HK$ band, we obtained a slightly lower $T_{\rm ex}$ with an either similar to or at most $\sim 2$ mag
higher $A_{V}$. We note that a few spectra from which we derived $A_{V} \sim 0$  yield 
the lowest $\chi^{2}$ for negative values of  $A_{V}$, which is unphysical. As sometimes
this occurs when the band is limited to $HK$ as well, which means that it is not always caused by the addition
of the $JH$ band, a possible explanation is that emission from higher vibrational levels originates
in warmer gas. A gas stratification with different temperatures has been noted
by \citet{2006A&A...449.1077C} for example. When we limit the fit to the $JH$ points, we obtain significantly higher
temperatures with an extinction either similar to or  different by about $\pm 2$ mag,
confirming that the emission from higher vibrational levels may arise from hotter gas.
By limiting the fit to the points in the KB grism band, we obtain slightly lower $T_{\rm ex}$
(by few 10 \%)
than when the whole $JHK$ band is used, usually (but not always)
with largely overestimated extinctions, provided that the vibrational 2--1 transitions
are also detected. Thus, the KB grism band only
is  not suitable to constrain the extinction. This is plausible because the KB grism band does not
contain pairs of transitions originating in the same upper level.

Another interesting feature is evident in Fig.~\ref{fig:KB}. Knots A15, F5,  and G3 exhibit a more intense
H$_2$ 1--0 S(3) line 
($1.9576$ $\mu$m) 
than the S(1) line. The knots in strand A display
S(3) lines that are even a factor of $>2$ brighter than the S(1) line after telluric correction. J-shock (e.g. \citealt{1995A&A...296..789S}) and fluorescent excitation \citep{1989ApJ...338..197S} models may  both exhibit H$_2$ 1--0 S(3)
lines that are slightly brighter than the S(1) transition, but possibly not such high ratios.
The telluric correction of H$_2$ 1--0 S(3) must therefore be investigated. 
This line falls inside a telluric CO$_{2}$ band (Fermi triad). Figure~\ref{fig_atmo} shows
that this band causes the largest absorption between $\sim 2.05-2.07$ $\mu$m,
far from the S(3) line.
However, this region is in the wing of the deep
water absorption separating the $H$ and $K$ bands, so it can be affected by large water vapour variations.
Table~\ref{KB:telcor} compares the S(3)/S(1) line ratios, both corrected and uncorrected for
telluric absorption, for all knots in each frame. For each frame, a comparison is also displayed
with the telluric star spectrum used for the correction. Although a difference in atmospheric water vapour
content between the telluric star and frame A observations can explain part of the line ratio, high line ratios are obtained
throughout the night, regardless of the telluric spectrum. In addition, \citet{2006A&A...449.1077C}  found
S(3)/S(1) line ratios of $\sim 2$ for 
knots A1--A15 
and $\sim 1.4$ for 
knots B1--B11 
(see their Table~22), as well as ratios $> 1$ in
several other regions.  The problem may be inherent in 
the telluric correction method, maybe related to the intrinsic absorption hydrogen Brackett line Br8 in the
telluric star spectrum near to the H$_2$ 1--0 S(3) line.
In this case, the corrected H$_2$ 1--0 S(3) line flux would be systematically
overestimated for all knots. Nevertheless, we verified that there is no significant difference in
telluric correction in the S(3) region between using the A0 or the G2 telluric star.

We note that line ratios from different knots in the same frame can be compared to each other regardless
of the telluric correction, that is, ratios normalised to that of a reference knot do not depend on atmospheric
correction.  
Knots A1--A15 
clearly display the highest S(3)/S(1) ratios. A check of the 2D spectra that were not corrected
for telluric extinction also confirms that the emission in a few of these knots is dominated by the
S(1) and S(3) lines in the KB grism band. If 
knots C1--C17 
are only slightly 
extincted
this may indicate different excitation or physical conditions for 
knots A1--A15. 
The $JH$ spectra of 
knots A1--A15  
do not exhibit H$_2$ lines in the $J$ band, maybe due to extinction, and the lines in the $H$ band cannot
be inter-calibrated with the lines in the KB grism band. The photo-ionised strip also exhibits a high
S(3)/S(1) ratio, which may be accounted for by the ongoing H$_2$ formation mechanism, as discussed above.

We detected [\ion{Fe}{II}] lines only towards knots A7, A11, A12, C14, and O1, confirming the results of \citet{2006A&A...449.1077C},
who clearly found [\ion{Fe}{II}] emission at $1.644$ $\mu$m from A1, A7, A11, A10--A12, and HH~593, but not
towards
knots C1--C17  
(see their Fig.~17). In addition, knots A7, A11, A12, and O1 do not exhibit H$_2$ lines
in the $J$ band.
Emission in the [\ion{Fe}{II}]  $1.644$ and $1.257$ $\mu$m, which arises from the same upper level,
is intense enough to allow us to derive the extinction
to those knots following the prescriptions by \citet{2016PASP..128g3001P}. For C14, both the $1.257$ and the $1.320$
$\mu$m lines fall in a region with relatively intense H$_2$ ro-vibrational lines, making it difficult even to
estimate upper limits. For the other knots, we found $A_{V}$ in the range 5--10 mag,
which is consistent with $A_{V} = 10 \pm 5$ quoted by \citet{2006A&A...449.1077C} for 
knots A1--A15  
and 
knots B1--B11. 
Thus, our detection
of H$_{2}$ lines in the $J$ band towards 
knots C1--C17  
and their lack towards 
knots A1--A15  
and 
knots B1--B11 
is consistent with the hypothesis that 
knots C1--C17 are less
extincted than 
knots A1--A15 and knots B1--B11. 
 \citet{2006A&A...449.1077C} reported, based on a sample of various star-forming regions including
some of knots A1--A15 and B1--B11, that bright H$_2$ lines in the $J$ band are generally observed
where $A_{V} < 5$ and undetected where $A_{V} > 10$. On the other hand, the [\ion{Fe}{II}] line
emission and the high S(3)/S(1) ratios towards the more extincted 
knots A1--A15
both clearly indicate different excitation mechanisms between 
knots C1--C17 and knots A1--A15
(C shocks versus\ J shocks?).

%
%
\begin{table}
\tiny
\centering
\caption{\label{tex:ext}
Excitation temperature, extinction, and reduced $\chi^{2}$
 from fits to the H$_2$ lines (plus $A_{V}$ from [\ion{Fe}{II}] line ratios).
The listed knots can be identified in Fig.~\protect\ref{find:chart}.
}
\begin{tabular}{llll}
\hline
knot($^{a}$) & $T_{\rm ex}$ & $A_{V}$ & Reduced\\
     & (K) & (mag) & $\chi^{2}$\\
\hline
\multicolumn{4}{c}{$HK+JH$ band($^{b}$)}\\
\hline
C6 & $2305 \pm 180$ & $5 \pm 3$ & $0.2$ \\
C7 & $2417 \pm 15$ & $0 \pm 0.3$ & 21 \\
C10 south & $2734 \pm 27$ & $0 \pm 0.3$ & 12 \\
C12 north & $2632 \pm 19$ & $0 \pm 0.3$ &  39 \\
C14 & $2266 \pm 10$ & $0 \pm 0.2$ & 146 \\
C16 north & $2336 \pm 31$ & $6.3 \pm 0.5$ & 35 \\
\hline
\multicolumn{4}{c}{$HK$ band($^{b}$)}\\
\hline
C17 north & $1292 \pm 239$ & $0 \pm 4$ & 1 \\
\hline
\multicolumn{4}{c}{$JH$ band($^{b}$)}\\
\hline
C15 & $3589 \pm 221$ & $1.2 \pm 0.7$ & 5 \\
A7$^{c}$ & -- & $9 \pm 3$ & -- \\
A11$^{c}$ & -- & $5 \pm 2$ & -- \\
A12$^{c}$ & -- & $7 \pm 2$ & -- \\
O1$^{c}$ & -- & $7 \pm 2$ & -- \\
\hline
\multicolumn{4}{c}{KB grism band($^{b,d}$)}\\
\hline
C8 & $2215 \pm 42$ & $14 \pm 1$ & 15 \\
C10 & $2246 \pm 29$ & $19 \pm 1$ & 11 \\
east of C8--C10 & $2278 \pm 89$ & $4.4 \pm 2.4$ & $0.2$ \\
north-east of C14 & $2182 \pm 59$ & $21 \pm 2$ & 5 \\
C14 & $2241 \pm 31$ & $14 \pm 1$ & 5 \\
A7$^{e}$ & $2089 \pm 124$ & $0 \pm 8$ & $0.4$ \\
A15$^{e}$ & $2425 \pm 137$ & $0 \pm 9$ & $0.05$ \\
B8 & $2617 \pm 285$ & $22 \pm 18$ & $0.3$ \\
G1$^{e}$ & $2240 \pm 275$ & $16 \pm 11$  & 3 \\
G2$^{e}$ & $2331 \pm 268$ & $7 \pm 12$  &  $0.2$ \\
G3$^{e}$ & $2249 \pm 72$ & $0 \pm 9$  &  $0.3$ \\
G4$^{e}$ & $2405 \pm 145$ & $5 \pm 11$  &  $0.4$ \\
\hline
\end{tabular}
\tablefoot{
\tablefoottext{a}
{Following the notation used in \citet{2009A&A...504...97B}.}
\tablefoottext{b}
{Spectra spanning the indicated band.}
\tablefoottext{c}
{$A_{V}$ from $1.644$/$1.257$ $\mu$m [\ion{Fe}{II}] line ratios.}
\tablefoottext{d}
{Fits in this band underestimate $T_{\rm ex}$ and overestimate $A_{V}$ (see text).}
\tablefoottext{e}
{H$_2$ 1--0 S(3) line point at $1.957$ $\mu$m not included in the fit to obtain a consistent
result.}
}
\end{table}

The excitation temperature and extinction derived from the fit to the H$_2$ lines and the
[\ion{Fe}{II}] line ratios are listed in Table~\ref{tex:ext}. We recall the caveats
noted above when only part of the $JHK$ band is available.
We also note that where the H$_2$ 1--0 S(3) line is much brighter than the H$_2$ 1--0 S(1),
this has not been used for the fit.

\subsection{H$_{2}$ luminosities}

After we derived 
a suitable range for extinction and excitation temperature, we used the H$_{2}$ $2.12$
$\mu$m line fluxes from \citet{2009A&A...504...97B} and the ratio of flux at $2.12$ $\mu$m to total H$_{2}$
ro-vibrational lines flux
computed by \citet{2006A&A...449.1077C} assuming LTE conditions to constrain the 
total H$_{2}$ luminosity 
of the
various knots. From Table~\ref{tex:ext}, we assumed $T_{\rm ex} = 2400 - 2700$ K and
$A_{V}$ in the range 0--2 mag for C, 5--10 mag for A and B, and  2--10 mag for G. The extinction
correction is based on the reddening law by \citet{1985ApJ...288..618R}. As for 
knots C1--C17, 
$A_{V} = 2$ appears as a
conservative upper limit for most of the knots, but to estimate the extinction range of G, we took into account
that the extinction is usually overestimated from fitting the H$_2$ lines in the KB grism band alone. 
The ranges of total H$_2$ luminosities, which also assume isotropic emission, are
listed in Table~\ref{lh2:lum}. The luminosities of the single knots provided in Table~\ref{lh2:single:lum} were summed
to derive the values in Table~\ref{lh2:lum}.

The total H$_2$ luminosity of 
knots C1--C17  
is clearly comparable to that of 
knots A1--A15 and B1--B11.
Using the
relation derived by \citet{2006A&A...449.1077C}, this luminosity could be associated with a driving source
of $L_{\rm bol} \sim 13 - 25$ $L_{\sun}$. This would indicate that the two driving sources in which knots C1--C17 originate,  
source  C and source I \citep{2012A&A...542L..26B}, are solar-mass protostars, or even sub-solar mass
protostars if the emission towards C is enhanced due to the outflow collision. This is also consistent with the
outflow mechanical luminosities derived from CO(1--0) by \citet{2012A&A...542L..26B}.
When the same relation is applied to this, the driving source producing strand G should have a bolometric luminosity
of $0.4$-2 $L_{\sun}$, that is, it would be a low-mass protostar.

%
%
\begin{table}
\tiny
\centering
\caption{\label{lh2:lum}
Total H$_2$ luminosity obtained assuming the extinction range indicated, $T_{\rm ex} = 2400 -2700$ K,
LTE conditions, and isotropic emission (scaled to a distance of 910 pc). 
The listed knots can be identified in Fig.~\protect\ref{find:chart}.}
\begin{tabular}{lll}
\hline
knot chain($^{a}$) & $L_{\rm H_2}$ & $A_{V}$ range \\
        &   ($L_{\sun}$) & (mag) \\
\hline
HH~593 & $(2.1\mbox{--}4.1) \times 10^{-2}$ & 5--10 \\
A     & $(1.5\mbox{--}2.8) \times 10^{-1}$ & 5--10 \\
B     & $(2.8\mbox{--}5.3) \times 10^{-1}$ & 5--10 \\
C     & $(1.8\mbox{--}2.6) \times 10^{-1}$ & 0--2\phantom{0}  \\
G     & $(2.4\mbox{--}6.2) \times 10^{-2}$ & 2--10 \\
\hline
\end{tabular}
\tablefoot{
\tablefoottext{a}
{Following the notation used in \citet{2009A&A...504...97B}.}
}
\end{table}

\section{Discussion}
\label{discu}

\subsection{Strands A and B}
The most remarkable result from the kinematical information (see Sect.~\ref{l:k:s}) are the nearby knots in
the  same strand with 
significantly different 
radial velocities (strands A, B, and G).
The simplest inference is that these jets lie roughly on the plane of the sky and do not
indicate an overlap of jets from different driving sources.

In this respect, the most notable example is probably knot A15, which resembles a bow shock whose
wings exhibit different radial velocities, as described above. We compared A15 with the 3D models of bow shocks
by \cite{2010A&A...513A...5G}. The knot spans $\sim 4\arcsec$ in width, which means $\sim 3600$ au at a distance
of $910$ pc. Assuming that each of the wings is $\sim 1\arcsec$ in radius, which is an upper limit, we
obtain from Table~5 that A15 north, for instance, has a mean brightness of $\sim 3.5 - 6 \times 10^{-7}$ W m$^{2}$ sr$^{-1}$
(which is a lower limit) after correcting for $A_V = 5 -10$. 
This is consistent with a bow shock seen almost edge-on that propagates in a medium with a density of $\sim 10^{4} - 10^{5}$
cm$^{-3}$ with a velocity of $40 - 60$ km s$^{-1}$. Both the excitation temperature ($\sim 2400$ K, see
Table~\ref{tex:ext}) and the 2--1 S(1)/1--0 S(1) ratio ($\sim 0.17$ in the north and $\sim 0.14$
in the south, see Table~3) are accounted for by the models as well. \cite{2010A&A...513A...5G} also predicted that
the wings of the bow shock exhibit a different radial velocity if the magnetic field does not lie in the plane of the sky. The larger difference arises when the projected magnetic field is roughly perpendicular to the direction
of propagation. The projected magnetic field derived by \cite{2018MNRAS.476.4782S} is even almost parallel
to the apparent axis of A15, but this holds for the edge of the cloud, so it is likely that the orientation inside
the cloud may be different. Most of the cases shown in \cite{2010A&A...513A...5G} adopt
a magnetic field strength of 500--1600 $\mu$G (in the density range of our interest), whereas \cite{2018MNRAS.476.4782S}
estimated a magnetic field of 220 $\mu$G towards \object{IC 1396N}. Again, this estimate holds at the tip of the cloud,
not in the inner regions. Thus  the main problem is that the velocity difference predicted for the cases
examined by \cite{2010A&A...513A...5G} is lower than 10 km s$^{-1}$, while we can deduce a difference of few
times 10 km s$^{-1}$ from Table~\ref{rad:vel}. In addition, the model parameter space should be explored in more
detail to determine not only whether this velocity difference can be matched, but also whether the structure
on a slightly larger scale,
with the knots trailing A15 north and A15 south exhibiting a similar velocity difference, can be reproduced. 
Interestingly, if A15 represents a real bow shock, then its axis points towards a position in
between \object{[BGE2002] BIMA 2} and \object{[BGE2002] BIMA 3}, rather than towards \object{[BGE2002] BIMA 2}, which is the driving source that is clearly indicated by the mm outflow. 

\begin{figure}
\resizebox{\hsize}{!}{\includegraphics{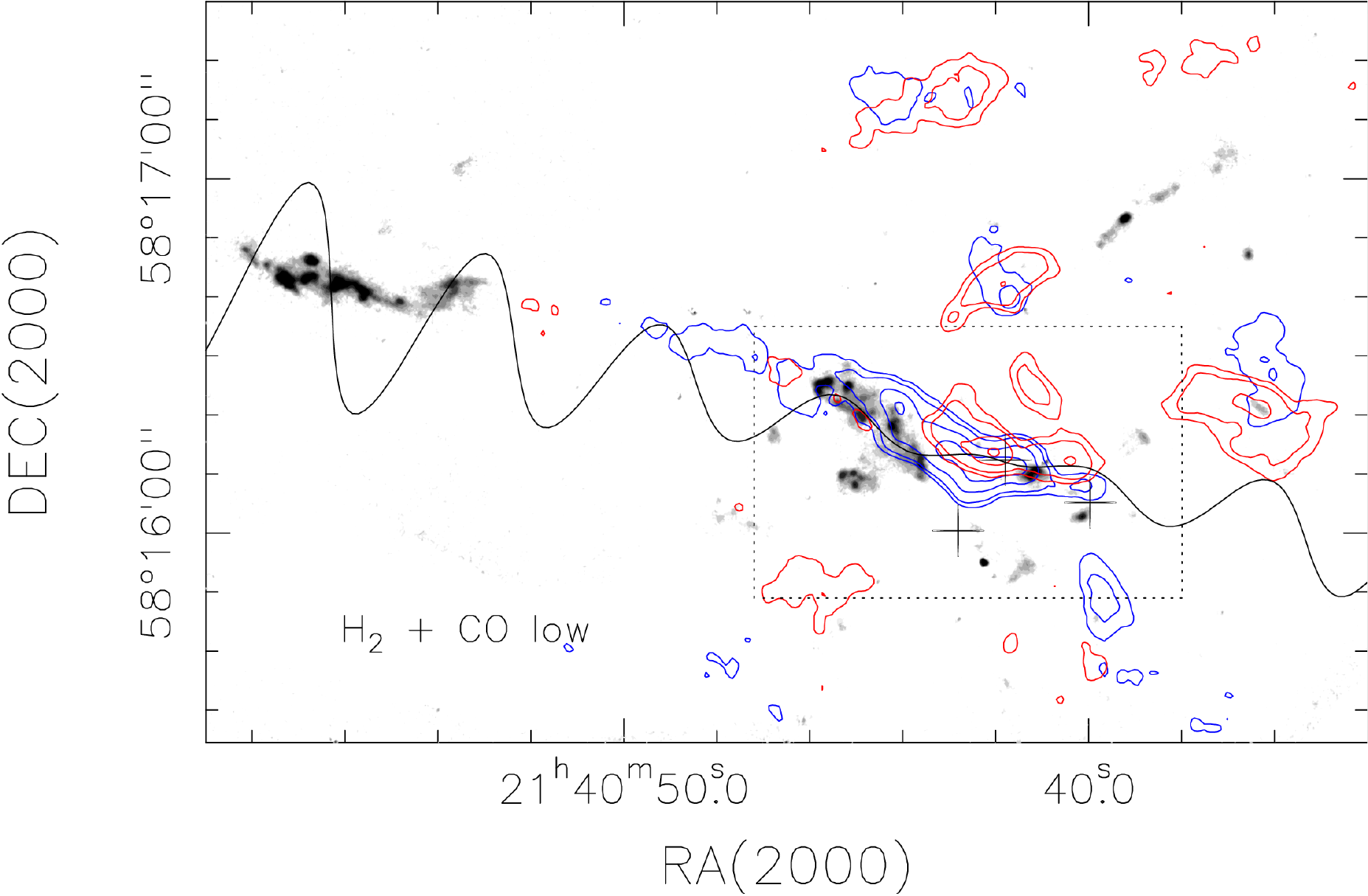}}
\caption{\label{fig:prec:jet}
Best fit of the projected morphology of a precessing jet as parametrised by \citet{1996AJ....112.2086E}
to the locations of the knots in strands A and B, overlaid with the H$_2$ $2.12$ $\mu$m image (continuum subtracted) 
in grey scale and CO ($J=1-0$) emission integrated in the [$\pm 3.5, \pm 9.5$] km s$^{-1}$
velocity interval in red and blue contours (adapted from Fig.~7a of \citealt{2009A&A...504...97B}).   
The crosses mark the location of the compact radio sources \object{[BGE2002] BIMA 1}, 2, and 3.
}
\end{figure}

The knots in the south-western part of strand A (A1, A4, A8, and A11) exhibit the same radial velocity
within the errors, a value between the extremes of A15 north and A15 south. When the systematic error 
is considered, this value is consistent with the velocity of the blue lobe of the outflow seen at mm
wavelengths that overlaps the strand (see 
Fig.~\ref{fig:prec:jet}, adapted from Fig.~7a of \citealt{2009A&A...504...97B}). 
This suggests that
these knots are associated with the same flow that 
causes
A15. [\ion{Fe}{II}] line emission has been observed
towards some of the knots, indicating the presence of J shocks in a few areas; in particular, 
\cite{2006A&A...449.1077C} also found emission ahead of A15 north, as expected if this bow-shock knot represents
the apex of the flow. Neither \cite{2006A&A...449.1077C} nor we detected [\ion{Fe}{II}] line emission towards
A15 south. 

An alternative explanation for the different radial velocities of A15 north and south is jet rotation.
The peaks of the two knots are $\sim 2\arcsec$ apart and the difference in radial velocity is $> 50$ km s$^{-1}$
(a lower limit that takes the systematic error into account and assumes that the outflow
lies on the plane of the sky). This would indicate a rotation velocity of $> 25$ km s$^{-1}$ at a distance
$\sim 1\arcsec$ (i.e. 910 au) from the outflow axis. Such a high rotational speed so far away from the driving source
would suggest that a rotational signature should be detected in the mm outflow near to the source as well.
Figure~\ref{fig:prec:jet} (or Fig.~7 of \citealt{2009A&A...504...97B}) indeed shows that blue-shifted and red-shifted mm emission
do not overlap near \object{[BGE2002] BIMA 2}. However, if this were the case,
rotation would occur in the opposite direction compared to 
what is derived from knot A15.  In addition, purely
hydrodynamic simulations of rotating jets do not reproduce the morphology of strand A and indicate that jet rotation
signatures are only preserved close to the driving source \citep{2007MNRAS.378..691S}.   
Since a new analysis of the old mm data is beyond the scope of this work, we just note that the available NIR
and mm observations both show how complex the scenario towards strand A is. Further high-resolution
observations are required.

Strand B is located east of strand A and is composed of a chain of knots that is roughly aligned in an east-west
direction. The radial velocity increases from west to east, ranging from $\sim -15$ km s$^{-1}$
(knot B3) to $\sim 11$ km s$^{-1}$ (knot B8). Even when the systematic error is taken into account, these
radial velocities are close to the systemic velocity, indicating that the knots move near to the
plane of the sky. This would confirm that strand B may be part of the same outflow system that causes
strand A, as suggested in the literature \citep{2009A&A...504...97B}. Other similarities between B and
A are the H$_2$ luminosity (B is roughly twice as luminous as A, see Table~\ref{lh2:lum}) and the extinction
range. Strand B exhibits a wiggled morphology that may result from jet precession, which would explain
the radial velocity variations along the strand as well. The strand is $\sim 1 \arcmin$ in length
(i.e. $\sim 0.26$ pc, or $\sim 54000$ au, at 910 pc) and assuming a projected velocity of 100 km s$^{-1}$ this results in
a precession period of $< 2600$ yr.

If strands A and B are caused by the same
jet (driven by \object{[BGE2002] BIMA 2}), then the jet is precessing. This link is in principle indicated by
the mm CO emission outlining the outflow, which stretches from A to B but does not overlap
B (see Fig.~\ref{fig:prec:jet}, adapted from Fig.~7a of \citealt{2009A&A...504...97B}).
To further study this scenario, we tried fitting
the function given by \citet{1996AJ....112.2086E} to the positions of the knots in strands A and B
from \citet{2009A&A...504...97B}. The free parameters in this simple model are the precession
amplitude $\alpha$, the precession length scale $\lambda$, the position angle $\psi$  of the outflow axis  
on the plane of the sky, and the initial phase $\phi_{0}$ at the driving source. As shown in Fig.~\ref{fig:prec:jet},
our best fit does not account simultaneously for the spatial arrangement of the two strands. We tried 
to constrain some of the free parameters as well, but were unable to find a better visual agreement. 
In Table~\ref{prec:eisl} we list the parameters obtained from the fits to strand A, to strand B, and to 
strand A and B simultaneously, all assuming \object{[BGE2002] BIMA 2} as the driving source. Clearly, if \object{[BGE2002] BIMA 2} did cause strands A and B, then a simple jet precession is not sufficient to explain 
their locations. One possible scenario is that an event between the ejection of B and A has resulted
in changing the jet position angle, increasing the amplitude and decreasing the timescale of the precession
(see Table~\ref{prec:eisl}). This would require a close encounter of \object{[BGE2002] BIMA 2} with another stellar 
object during an accretion phase \citep{2000MNRAS.317..773B}, for example. Another scenario is that strand A is caused by the interaction of two or more jets from different driving sources. Finally, A and B could have
been caused by two different driving sources. At present, our data do not allow us to distinguish between
these three possible interpretations.

%
%
%
%

%
%
\begin{table}
\tiny
\centering
\caption{\label{prec:eisl}
Jet parameters obtained from fitting the function by \protect\citet{1996AJ....112.2086E} to
the locations of 
knots A1--A15 and/or B1--B11.
We always assumed \object{[BGE2002] BIMA 2} as the driving source.
}
\begin{tabular}{lllllll}
\hline
Strand fit & $\chi^{2}$ & $\psi$ & $\alpha$ & $\lambda$ & $\chi_{0}$ & T($^{a}$) \\
           &  & $(\degr)$ & $(\degr)$ & $(\arcsec)$ & $(\degr)$ & (yr) \\ 
\hline
A & $0.15$ & $-82$ & 18 & 44 & 35 & 1899 \\
B & $2 \times 10^{-3}$ & $-71$ & 4 & 128 & 108 & 5525 \\
A and B & $0.32$ & $-77$ & 9 & 32 & $-51$ & 1380 \\ 
\hline
\end{tabular}
\tablefoot{
\tablefoottext{a}
{Obtained from $\lambda$ by assuming a jet velocity of 100 km s$^{-1}$.}
}
\end{table}

\subsection{Other emission features}
The chain of knots G exhibits a radial velocity pattern similar to that of strand B. The knots
are aligned in a south-east to north-west direction, and the radial velocity increases from
knot G1 ($-7$ km s$^{-1}$) to G3 (40 km s$^{-1}$), then decreases again to G7 ($-31$ km s$^{-1}$).
These values oscillate around a value close to the systemic velocity
(i.e. $\sim 0$ km s$^{-1}$), even when 
the systematic errors are taken into account, suggesting that in this case, the outflow also lies close to the
plane of the sky. 
The most likely explanation is jet precession or the orbital motion
of the driving source in a double system (e.g. \citealt{2012AJ....144...61E}). 
If the knots are much closer
than strand B to the driving source, we can expect that a wiggled morphology has not yet developed 
above the spatial resolution of the available images. 
Using the same computation as for B, we obtain a precession period of $\sim 1300$ yr.
Strand G has commonly been associated with the infrared source HH 777/IRS331, which also seems to drive
a perpendicular outflow originating the Herbig-Haro object HH 777 \citep{2009A&A...504...97B}.
If this is the case, HH 777/IRS331 would be a double system, and jet precession or the effects of orbital motion are both plausible.

Strand F is associated with the red lobe of the outflow from source I \citep{2012A&A...542L..26B},
and its knots are therefore expected to exhibit radial velocities in the range $\sim 0-20$ km s$^{-1}$
(see Fig.~2 of \citealt{2012A&A...542L..26B}). Although the radial velocity of knots F1 and F5
agrees, the knots in between, F2 and F4, are clearly blue-shifted with radial velocities
of $-11$ and $-33$ km s$^{-1}$, respectively. On the other hand, 
knots C1--C17   are associated with the
blue lobes of the outflows from sources I and C and display radial velocities in the range
$-8$ to $-47$ km s$^{-1}$, in agreement with the radial velocities obtained from the
mm emission. The pattern in the radial velocities of 
knots F1--F5 
is not easily interpreted
as jet precession. Alternatively, it could be due to oblique shocks developing due to gas inhomogeneities. 
The detailed pattern exhibited by 
knots C1--C17   
is more difficult to interpret as this region results from
the overlap (possibly the collision) of the two different outflows. This pattern is explored
in detail in a twin paper \citep{2022A&A...661A.106L} using proper motion determination. 
 
We note that the difference (i.e. not the single values) in radial velocities between knots
{\it \textup{in the same strand}} never exceeds the critical speed for H$_2$ dissociation (e.g.
\citealt{2010A&A...513A...5G}), which confirms that each strand represents one single system and did not
originate from the overlap of different jets.  

The most notable result from the excitation analysis of the H$_2$ emission from the knots in strand C
is that they appear to be less extincted than the other features in the cloud. This suggests that they
are closer to the cloud surface facing the observer. We detected [\ion{Fe}{II}] emission only towards knot C14,
the brightest in the strand (see Table~5), although we would expect more widespread [\ion{Fe}{II}] emission if the outflow-collision
scenario were correct. In this respect, a clue might come from the presence of higher temperature gas, which 
unfortunately is only hinted at by our data. 

The only other knot in which we detected [\ion{Fe}{II}] emission is O1, which also exhibits a high H$_2$ 1--0 S(3)/S(1) 
flux ratio, as found towards A. It is a highly reddened ($A_V \sim 7$) red-shifted ($\sim 22$ km s$^{-1}$)
isolated feature aligned with strand F. This is consistent with O1 being a terminal shock associated with the
red lobe of the outflow driven by source I.

The photo-ionised strip borders the south-western rim of the dark optical nebula (see e.g. Fig.~1 of
\citealt{2001A&A...376..271C}) and is roughly perpendicular to the direction to HD20626. A patch of blue-shifted
molecular gas was observed towards this part of the dark nebula (see Fig.~3 of
\citealt{2001A&A...376..271C}); its radial velocity, $\sim -5$ km s$^{-1}$, coincides with the radial velocity 
we measured for the strip (see Table~\ref{rad:vel}) within the errors. Its H$_2$ emission therefore marks 
the PDR facing the ionising source, seen edge-on. Its blue-shifted radial velocity is likely to be the signature of
a photo-evaporation process. However, a PDR should expand at the sound speed of the warm gas (a few km s$^{-1}$;
\citealt{2002ApJ...573..215G}), hence the H$_2$ radial velocity we found, if confirmed, cannot be
accounted for by gas expansion alone.
 
\subsection{Jets from sources in orbital motion} 

An alternative to jet precession is that the driving source is in orbital motion in a
double system, as we described above. Although in some cases the two scenarios may be hardly distinguishable 
(e.g. see \citealt{2022A&A...657A.110F}), we can show that this can probably be ruled out for strand A.
Based on the simple formulation by \citet{2002ApJ...568..733M}, if the driving source is located
towards \object{[BGE2002] BIMA 2}, we can estimate the orbital
radius $r_{0}$ from their Eq.~13 (under the assumption of a circular orbit) deriving $\Delta z$ from
our Fig.~\ref{fig:prec:jet} ($\sim 35\arcsec$) and a lower limit for $k = \tan \alpha$ from
Table~\ref{prec:eisl} ($\alpha \sim 9 \degr$). Then, we can estimate a minimum mass $M_{\rm p}$ for the
driving source from their Eq.~1, using a period $T \sim 2000$ yr
from Table~\ref{prec:eisl} and noting that $M_{\rm p}$ is minimum when $M_{\rm p}/M_{\rm s} = 1$
($M_{\rm p} \geq M_{\rm s}$). We obtain $M_{\rm p} \sog 500$ $M_{\sun}$, which is clearly
inconsistent. We conclude that the morphology of strand A cannot be produced by a driving source that coincides
with \object{[BGE2002] BIMA 2} and is in orbital motion. 

\section{Summary and conclusions}
\label{conclu} 

We have presented long-slit NIR $J$, $H$, and $K$ spectroscopy of the \hdos{} outflows reported by 
\citet{2009A&A...504...97B} in the bright-rimmed cloud \object{IC 1396N}.
The analysis of the spectra allowed us to derive physical properties (from the grisms spanning the $JH$
and $HK$ bands with $R \sim 500$ and the grism spanning the $K$ band with $R \sim 1200$)
and kinematics (from the grism spanning the $K$ band with $R \sim 1200$) for the  
knots encompassed by the slit (using several pointings and position angles). The main results are listed below.
\begin{enumerate}

\item
The spectra of the jet knots show a set of ro-vibrational lines of H$_2$ and \fe\ forbidden lines.
The \fe\ lines are only detected towards
knots A7, A11, A12, C14, and O1.

\item
We derived the radial velocity of the jet knots from the  H$_2$ 1--0 S(1) transition.
In addition, we verified that the radial velocities derived from the S(0) and S(2) 
transitions (where these lines are detected with a good S/N)
are consistent  within the errors with the S(1) results.

\item
We found blue-shifted and red-shifted radial velocity values ranging from $-40$ to $+50$ km s$^{-1}$.
For the strands of knots A and C, the blue-shifted and red-shifted emission is  fully consistent with the mm data from the CO emission of
\citet{2009A&A...504...97B} and \citet{2012A&A...542L..26B}.

\item
Knots F1--F5 
lie in the red-shifted lobe of one of the CO outflow.
We found red-shifted velocities for knots F1 and F5, in agreement with CO data, while F2 and F4 have blue-shifted velocities.
Knots B1--B11 and G1--G7 
show two separate chains of blue-shifted and  red-shifted knots.

\item
The simultaneous presence of blue-shifted and red-shifted knots in the same strands indicates
that the outflows roughly lie on the plane of the sky. It also suggests that jet precession is ubiquitous
in the region, as was confirmed by the wiggling morphology of strand B, for example. 

\item
A photo-ionised strip, located south-west of strand A, marks the edge-on PDR at the side of the cloud facing the ionising star and  
displays a blue-shifted velocity, which may (at least partly) be accounted for by gas photo-evaporation. 

\item
The H$_2$  line ratios 1--0~S(1) to 2--1~S(1), and 1--0~S(1) to 1--0~S(0) were used to distinguish between shock-excited and
UV-fluorescent mechanisms giving rise to the H$_2$ emission.
The ratios we found for the knots of the chains we mapped are consistent with shock excitation as expected for the emission of 
knots associated with H$_2$ outflows.
These line ratios are significantly different for the photo-ionised strip south-west of the chain of knots  A, 
which confirms a fluorescent mechanism of the emission.

\item
We used all the H$_2$ detected lines to simultaneously derive the excitation temperature ($T_{\rm ex}$) and the 
extinction ($A_V$) assuming that the gas is thermalised.
The $T_{\rm ex}$ obtained (2000--3000 K) is typical of H$_2$ jet knots from Class~0 and Class~I sources.
We found that the chain of  knots C has a lower extinction ($A_V$ in the range of 0--2 mag) than 
chains A, B (5--10 mag), and G (2--10 mag).
This is consistent with the lack of detection of H$_2$ lines in the $J$ band toward 
knots A1--A15 and B1--B11, 
while these lines are detected in 
knots C1--C17.
In addition, the higher-excitation lines indicate higher-temperature layers of gas.

\item
The detection of \fe\ lines and the high H$_2$ 1--0 S(3)/S(1) ratios towards the more extincted knots of the A chain and O1
indicate
that different excitation mechanisms (C shocks versus J shocks) cause the emission from 
knots C1--C17, O1, and A1--A15.
However,
because the H$_2$ 1--0 S(3) wavelength coincides with
a spectral region in which atmospheric extinction is an issue, the ratio
H$_2$ 1--0 S(3)/S(1) in these (and similar other features in a number of regions) 
deserves further investigation. Our data strongly suggest that
the high values of this ratio are real.

\item
We derived the total H$_2$ luminosity of the knots, assuming LTE conditions and the ranges of $A_V$ and $T_{\rm ex}$ found previously.
The luminosities of strand C are consistent with the hypothesis that the mm sources I and C of \citet{2009A&A...504...97B} 
are solar or sub-solar mass (proto)stars.

\item
We found that a simple precession model cannot account for the hypothesis that strands A and  B both originate from source \object{[BGE2002] BIMA 2},
as suggested in the literature. 
We also ruled out that the outflow morphology results from a driving source in orbital motion towards \object{[BGE2002] BIMA 2},
which would need to possess an inconsistently high mass.
This indicates either a more complex scenario or that these strands are not
produced by the same driving source.

\item
By comparing strand A with numerical 3D models, we determined that the eastern tip of the strand is
consistent with a bow shock leading an outflowing parcel of gas. 
Knot O1 is
likely to represent a terminal shock associated with strand F and the red lobe of the outflow from
mm source I.

\end{enumerate}

Further high spatial-resolution mm observations are needed to study the relation (if any) between
strands A and B and find out the exact origin of the radial velocity oscillations displayed within
some of the strands.  

\begin{acknowledgements}

This work has been partially supported by the Spanish MINECO grants
AYA2014-57369-C3 and AYA2017-84390-C2 (co-funded with FEDER funds) and
MDM-2014-0369 of ICCUB (Unidad de Excelencia `Mar\'{\i}a de Maeztu').
We would like to thank the referee, A. Raga, for his insightful comments.

\end{acknowledgements}

%
%

\bibliography{ic1396_bib}

\begin{appendix}

\section{Atmospheric transmission}

\begin{figure}[h]
\resizebox{\hsize}{!}{\includegraphics{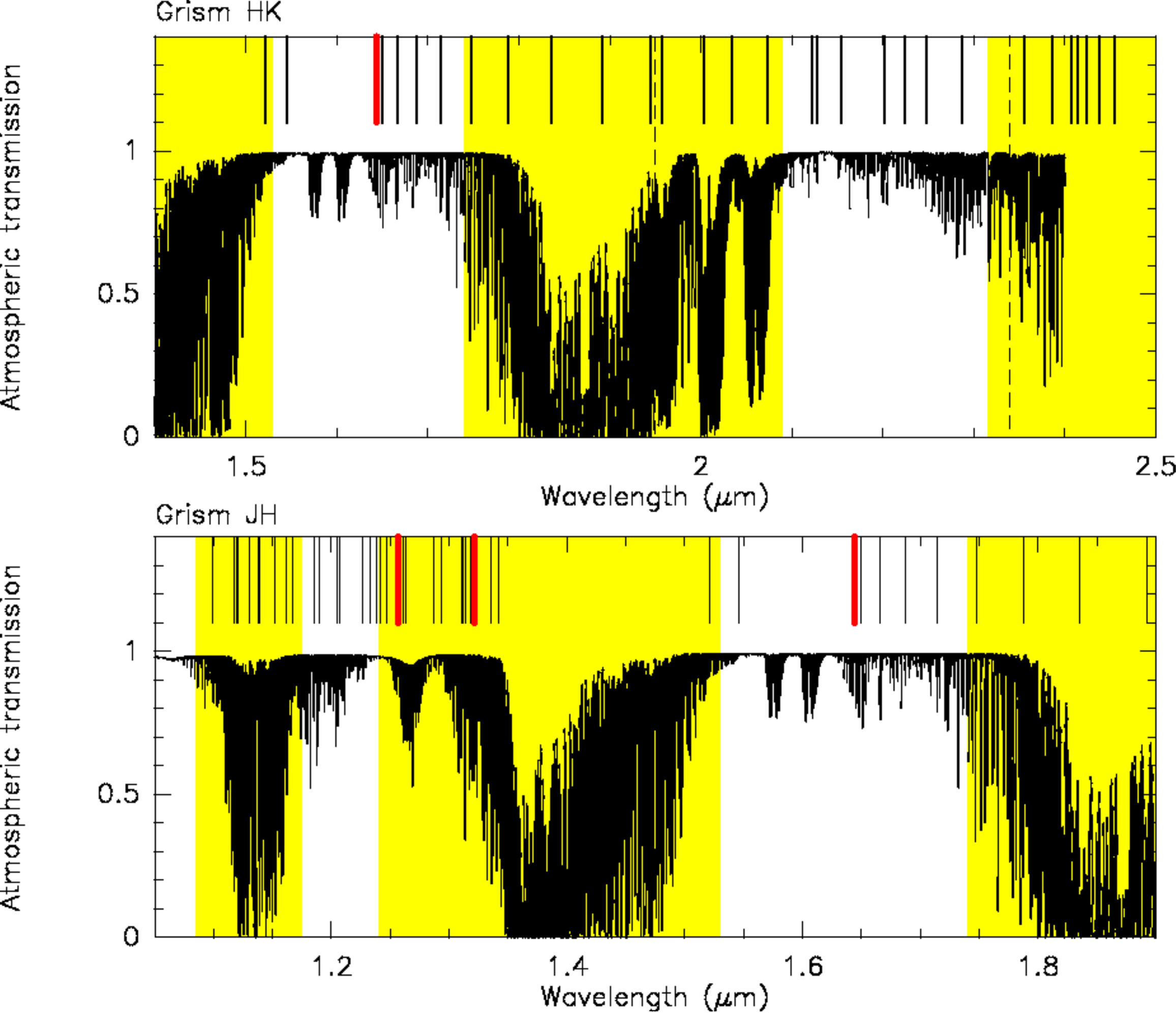}}
\caption{\label{fig_atmo}
Atmospheric transmission in the bands spanned by grism HK (top panel) and
JH (bottom panel). The vertical bars indicate the wavelengths of the H$_{2}$
lines (black) and [\ion{Fe}{II}] lines (red) discussed in the main text. The yellow areas
mark the wavelength ranges within which correction for atmospheric extinction
may exhibit large errors. The range spanned by grism KB is limited by the vertical dashed
lines.  }
\end{figure}

We used the {\it ESO Advanced Cerro Paranal Sky 
Model\footnote{https://www.eso.org/sci/software/pipelines/skytools/skymodel}}
\citep{2012A&A...543A..92N} to build a picture of the atmospheric transmission
in the $JHK$ bands and identify the more critical wavelength ranges for spectroscopy
(Fig.~\ref{fig_atmo}). Lines falling in the wings of the water telluric absorption features
separating the $JHK$ bands clearly need to be considered with caution, especially because
the transmission in these regions can be highly variable. 

\FloatBarrier

\section{Telluric correction}
\label{tel:cor:ap}

We performed a simple
test of the accuracy of our telluric correction  by dividing pairs of 1D spectra of the same 
telluric standard star taken at different times during the same
night. This clearly
removes the intrinsic stellar spectrum, and the obtained curve, 
$r(\lambda)\, [=T_{1}(\lambda)/T_{2}(\lambda)]$, allows mapping the differences
in flux ratios between two wavelengths when
the telluric correction curve derived from one of the two stellar spectra rather than the other is used. In particular,
if a wavelength $\lambda_{\rm a}$ is selected as a reference (we used $2.1$ $\mu$m for grisms KB and HK, and $1.7$
$\mu$m for grism JH), we can derive the difference in the flux ratio
$R_{i}(\lambda_{\rm a},\lambda_{\rm b})\, [=S(\lambda_{\rm b})/S(\lambda_{\rm a})]$ of two hypothetical lines centred at $\lambda_{\rm a}$ and
$\lambda_{\rm b}$ when using either the first ($R_{1}$) or the second
($R_{2}$) stellar spectrum to correct for a target spectrum. The quantity $r(\lambda)$
 typically has an
average value $\sim 1$ and exhibits small oscillations or even a slight gradient. The average value is not relevant for the line ratios because its
variations are due to seeing changes, pointing inaccuracies, different airmasses, and so on during the two integrations
on the telluric stars.
The average value does not affect the line ratios because\begin{equation}
\frac{r(\lambda_{\rm a})}{r(\lambda_{\rm b})} = \frac{R_{1}(\lambda_{\rm a},\lambda_{\rm b})}{R_{2}(\lambda_{\rm a},\lambda_{\rm b})}
\end{equation}
yields the relative difference in the ratios of flux at $\lambda_{\rm b}$ and at our reference wavelength
$\lambda_{\rm a}$ between the two corrections.

Standard spectral ratios $r$ between pairs
of stellar spectra in the same ABBA cycle indicate flux ratio changes of at most $\sim 5$ \% in the HK and KB 
grism bands,
and $\sim 10$ \% in the $JH$ band. Thus, we averaged all four 1D spectra from any ABBA cycle before constructing the
associated telluric correction curve
(Sect.~\ref{ODR}). No significant difference is found between the correction spectra obtained
from the A0V and the G2V stars. Finally, even the ratios $r$ obtained from the average spectra taken at
the beginning, in the middle and at the end of a night, point to maximum line ratio differences of $\sim 10$ \%
when both lines fall outside the regions of strong telluric absorption (see Fig.~\ref{fig_atmo}),
particularly water.

The telluric lines or bands at $\sim 1.24 - 1.28$ $\mu$m (O$_2$), $\sim 1.56 - 1.6$ $\mu$m (CO$_2$ and O$_2$),
$\sim 2 - 2.08$ $\mu$m (CO$_2$), $\sim 1.65  - 1.7$ $\mu$m and $\sim 2.24 - 2.3$ $\mu$m (CH$_4$) showed variations during the same night  that are mostly related with the airmass, as expected. The maximum flux ratio changes
were found
for $\lambda_{\rm b}$ coinciding with the wavelength of the peak of the strongest CO$_2$ absorption line at $\sim 2$ $\mu$m
($\sim 30$ \% for an airmass difference of $\sim 0.4$) and for $\lambda_{\rm b}$ inside the range of the O$_2$
band at $\sim 1.24 - 1.28$ $\mu$m ($\sim 25$ \% for an airmass difference of $\sim 0.4$).

\section{Telluric correction at the edge of the KB grism band}

A puzzling result is the high value of the ratio H$_2$ 1--0 S(3)/S(1) measured
towards some of the knots. Table~\ref{KB:telcor} compares the ratios before and
after telluric correction for every knot in each 
KB grism frame and lists their relation with the telluric standard star we used.
\FloatBarrier
%
%
\begin{table}[h]
\caption{\label{KB:telcor}
Details of the telluric correction of the H$_2$ 1--0 S(3) line in the KB grism band.
The listed knots can be identified in Fig.~\protect\ref{find:chart}.
}
\resizebox{7.1cm}{!}{%
\tiny
\begin{tabular}{lll}
\hline
knot($^{a}$) & Uncorrected & Corrected($^{b}$) \\
             & 1--0 S(3)/S(1) & 1--0 S(3)/S(1) \\
             & line ratio & line ratio \\
\hline
\multicolumn{3}{l}{\tiny frame A($^{c}$), exposure start UT 21:54, airmass $1.32$ position angle $52\fdg5$}\\
\multicolumn{3}{l}{\tiny telluric standard \object{HIP 109079},
   exposure start UT 20:54, airmass $1.65$ position angle $0\degr$}\\
\hline
A1 & $0.26$ & $1.7$ \\
A4 & $0.26$ & $1.6$ \\
A8 & $0.30$ & $2.2$ \\
A11 & $0.4$ & $3.1$ \\
A14 & $0.26$ & $1.7$ \\
A15 south & $0.41$ & $2.7$ \\
A16 & $0.65$ & $3.9$ \\
photo-ionised strip & $0.45$ & $2.6$ \\
\hline
\multicolumn{3}{l}{\tiny frame B($^{c}$), exposure start UT 22:34, airmass $1.25$ position angle $80\fdg5$}\\
\multicolumn{3}{l}{\tiny telluric standard \object{HIP 109079},
   exposure start UT 00:00, airmass $1.07$ position angle $55\degr$}\\
\hline
A12 & $0.59$ & $2.4$ \\
A15 north & $0.36$ & $1.53$ \\
B7 & $0.18$ & $0.95$ \\
B8 & $0.16$ & $0.78$ \\
\hline
\multicolumn{3}{l}{\tiny frame C($^{c}$), exposure start UT 23:14, airmass $1.19$ position angle $94\fdg5$}\\
\multicolumn{3}{l}{\tiny telluric standard \object{HIP 109079},
   exposure start UT 00:00, airmass $1.07$ position angle $55\degr$}\\
\hline
C6 & $0.26$ & $1.14$ \\
C8 & $0.18$ & $0.8$ \\
unclassified east of C8/C10 & $0.24$ & $0.96$ \\
C10 & $0.18$ & $0.73$ \\
C14 & $0.18$ & $0.77$ \\
unclassified north-east of C14 & $0.15$ & $0.67$ \\
C16 & $0.07$ & $0.30$ \\
\hline
\multicolumn{3}{l}{\tiny frame F($^{c}$), exposure start UT 02:39, airmass $1.24$ position angle $125\degr$}\\
\multicolumn{3}{l}{\tiny telluric standard \object{HIP 109079},
   exposure start UT 04:30, airmass $1.34$ position angle $100\degr$}\\
\hline
C1 & $0.19$ & $< 1.3$ \\
F1 & $0.4$ & $1.9$ \\
F2 & $0.37$ & $1.6$ \\
F4 & $0.26$ & $1.3$ \\
F5 & $0.32$ & $1.5$ \\
O1 & $0.55$ & $2.6$ \\
\hline
\multicolumn{3}{l}{\tiny frame G($^{c}$), exposure start UT 03:07, airmass $1.30$ position angle $130\degr$}\\
\multicolumn{3}{l}{\tiny telluric standard \object{HIP 109079},
   exposure start UT 04:30, airmass $1.34$ position angle $100\degr$}\\
\hline
A6/A7 & $0.32$ & $1.56$ \\
plateau A7 & $0.5$ & $2.3$ \\
G1 & $0.35$ & $1.7$ \\
G2 & $0.31$ & $1.3$ \\
G3 & $0.43$ & $1.9$ \\
G4 & $0.41$ & $2.0$ \\
G7 & $0.26$ & $1.2$ \\
\hline
\multicolumn{3}{l}{{\bf Notes} $^{(a)}$ Following the notation used in \protect\citet{2009A&A...504...97B}. $^{(b)}$ Corrected for}\\
\multicolumn{3}{l}{telluric absorption using for each frame the standard star observation summarised}\\
\multicolumn{3}{l}{below the indicated frame name. $^{(c)}$ The knots listed below have been recorded}\\ 
\multicolumn{3}{l}{simultaneously in the same frame.}
\end{tabular}
}
\end{table}

\section{L$_{\rm H_2}$ of single knots}

Table~\ref{lh2:single:lum} lists the total luminosity in H$_2$ ro-vibrational transitions
of single knots derived from
the photometry of H$_2$ 1--0 S(1) line emission at $2.12$ $\mu$m given by \citet{2009A&A...504...97B}.
The computation assumes the $T_{\rm ex}$ and $A_{V}$ ranges deduced in Sect.~\ref{li:ra:ds}
through linear fits  in the Boltzmann  plots to the detected 
H$_2$ lines or through  [\ion{Fe}{II}] line ratios,
the extinction law of \citet{1985ApJ...288..618R}, LTE conditions, and isotropic emission from the knots.
We used the relation between total H$_2$ and H$_2$ 1--0 S(1) fluxes in LTE as a function of
$T_{\rm ex}$ computed by \citet{2006A&A...449.1077C}.

\FloatBarrier
%
%
%
\begin{table}
\centering
\caption{\label{lh2:single:lum}
Range of total H$_2$ luminosities obtained assuming the extinction range indicated, 
$T_{\rm ex} = 2400 - 2700$ K, LTE conditions, and isotropic emission (scaled to a distance of 910 pc).
The listed knots can be identified in Fig.~\protect\ref{find:chart}.} 
\resizebox{7.1cm}{!}{%
\tiny
\begin{tabular}{lr@{$\,\pm\,$}lc@{\extracolsep{2em}}r@{\extracolsep{0em}$\,\pm\,$}lc}
\hline
& \multicolumn{3}{c}{Lower limit}
& \multicolumn{3}{c}{Upper limit}\\
\cline{2-4}\cline{5-7}
& \multicolumn{2}{c}{$L_\mathrm{H_2}$} & $A_V$   
& \multicolumn{2}{c}{$L_\mathrm{H_2}$} & $A_V$ \\
Knot(\tablefootmark{a}) 
& \multicolumn{2}{c}{($10^{-3} L_\sun$)} & (mag) 
& \multicolumn{2}{c}{($10^{-3} L_\sun$)} & (mag)   \\
\hline
HH~593A & $5.02$  & $0.07$ & 5 & $11.5$  & $0.1$  & 10 \\
HH~593B & $6.09$  & $0.09$ & 5 & $11.9$  & $0.1$  & 10 \\
HH~593C & $4.36$  & $0.07$ & 5 & $8.5$   & $0.1$  & 10 \\
HH~593D & $6.04$  & $0.10$ & 5 & $11.8$  & $0.1$  & 10 \\
A1     & $13.40$ & $0.15$ & 5 & $26.4$  & $0.3$  & 10 \\
A2     & $3.97$  & $0.12$ & 5 & $7.8$   & $0.1$  & 10 \\
A3     & $4.78$  & $0.09$ & 5 & $9.4$   & $0.1$  & 10 \\
A4     & $9.1$   & $0.1$  & 5 & $18.0$  & $0.3$  & 10 \\
A5     & $7.2$   & $0.1$  & 5 & $14.0$  & $0.3$  & 10 \\
A6     & $3.9$   & $0.1$  & 5 & $7.5$   & $0.3$  & 10 \\
A7     & $17.7$  & $0.1$  & 5 & $34.7$  & $0.3$  & 10 \\
A8     & $3.9$   & $0.1$  & 5 & $7.7$   & $0.3$  & 10 \\
A9     & $5.0$   & $0.1$  & 5 & $9.7$   & $0.3$  & 10 \\
A10    & $9.7$   & $0.1$  & 5 & $19.0$  & $0.3$  & 10 \\
A11    & $14.1$  & $0.3$  & 5 & $27.5$  & $0.4$  & 10 \\
A12    & $14.1$  & $0.1$  & 5 & $27.5$  & $0.3$  & 10 \\
A13    & $3.30$  & $0.09$ & 5 & $6.5$   & $0.1$  & 10 \\
A14    & $7.9$   & $0.1$  & 5 & $15.5$  & $0.3$  & 10 \\
A15    & $28.0$  & $0.1$  & 5 & $54.6$  & $0.3$  & 10 \\
B1     & $13.0$  & $0.3$  & 5 & $25.5$  & $0.4$  & 10 \\
B2     & $37.8$  & $0.3$  & 5 & $74.2$  & $0.6$  & 10 \\
B3     & $12.2$  & $0.1$  & 5 & $24.0$  & $0.3$  & 10 \\
B4     & $36.2$  & $0.3$  & 5 & $71.0$  & $0.4$  & 10 \\
B5     & $55.9$  & $0.3$  & 5 & $109.5$ & $0.4$  & 10 \\
B6     & $19.4$  & $0.1$  & 5 & $38.1$  & $0.3$  & 10 \\
B7     & $28.0$  & $0.1$  & 5 & $54.8$  & $0.3$  & 10 \\
B8     & $53.7$  & $0.3$  & 5 & $105.4$ & $0.4$  & 10 \\
B9     & $5.0$   & $0.1$  & 5 & $9.7$   & $0.1$  & 10 \\
B10    & $5.9$   & $0.1$  & 5 & $11.6$  & $0.3$  & 10 \\
B11    & $5.8$   & $0.1$  & 5 & $11.3$  & $0.1$  & 10 \\
C1     & $0.91$  & $0.07$ & 0 & $1.3$   & $0.1$  & 2  \\
C2     & $1.74$  & $0.04$ & 0 & $2.50$  & $0.06$ & 2  \\
C3     & $1.47$  & $0.04$ & 0 & $2.12$  & $0.07$ & 2  \\
C4     & $1.65$  & $0.06$ & 0 & $2.34$  & $0.07$ & 2  \\
C5     & $7.26$  & $0.04$ & 0 & $10.41$ & $0.06$ & 2  \\
C6     & $13.57$ & $0.04$ & 0 & $19.49$ & $0.06$ & 2  \\
C7     & $18.21$ & $0.06$ & 0 & $26.13$ & $0.07$ & 2  \\
C8     & $18.42$ & $0.06$ & 0 & $26.46$ & $0.09$ & 2  \\
C9     & $5.02$  & $0.06$ & 0 & $7.23$  & $0.09$ & 2  \\
C10    & $14.84$ & $0.07$ & 0 & $21.30$ & $0.09$ & 2  \\
C11    & $15.18$ & $0.07$ & 0 & $21.8$  & $0.1$  & 2  \\
C12    & $10.73$ & $0.06$ & 0 & $15.40$ & $0.07$ & 2  \\
C13    & $16.77$ & $0.07$ & 0 & $24.06$ & $0.09$ & 2  \\
C14    & $26.65$ & $0.06$ & 0 & $38.29$ & $0.07$ & 2  \\
C15    & $4.95$  & $0.09$ & 0 & $7.1$   & $0.1$  & 2  \\
C16    & $19.26$ & $0.07$ & 0 & $27.7$  & $0.1$  & 2  \\
C17    & $2.9$   & $0.4$  & 0 & $4.1$   & $0.6$  & 2  \\
G1     & $1.24$  & $0.03$ & 2 & $3.3$   & $0.1$  & 10 \\
G2     & $2.91$  & $0.04$ & 2 & $7.8$   & $0.1$  & 10 \\
G3     & $9.01$  & $0.07$ & 2 & $24.1$  & $0.1$  & 10 \\
G4     & $3.37$  & $0.06$ & 2 & $9.0$   & $0.1$  & 10 \\
G5     & $1.09$  & $0.03$ & 2 & $2.94$  & $0.09$ & 10 \\
G6     & $2.00$  & $0.06$ & 2 & $5.3$   & $0.1$  & 10 \\
G7     & $3.74$  & $0.09$ & 2 & $10.0$  & $0.1$  & 10 \\
O1     & $5.78$  & $0.1$  & $7\pm 2$ & \multicolumn{2}{c}{\ldots} & \ldots \\
\hline
\multicolumn{7}{l}{{\bf Notes.} $^{(a)}$ Following the notation used in \protect\citet{2009A&A...504...97B}.}
\end{tabular}
}
\end{table}

\end{appendix}

\end{document}